\documentclass{article}

\PassOptionsToPackage{sort&compress}{natbib}
\usepackage[final]{neurips_2025}
\usepackage[utf8]{inputenc} 
\usepackage[T1]{fontenc}    
\usepackage{hyperref}       
\usepackage{url}            
\usepackage{booktabs}       
\usepackage{amsfonts}       
\usepackage{nicefrac}       
\usepackage{microtype}      
\usepackage{xcolor}         
\usepackage{amsmath}
\usepackage{algorithm}
\usepackage{algpseudocode}
\usepackage{makecell}
\usepackage{arydshln}
\usepackage{array} 
\usepackage{multirow} 
\usepackage{amssymb} 
\usepackage{graphicx}
\usepackage{subcaption}

\algnewcommand{\LeftComment}[1]{\Statex \(\triangleright\) #1}

\title{Understanding protein function with a multimodal retrieval-augmented foundation model}

\author{
  Timothy F. Truong Jr \\
  OpenProtein.AI\\
  NY, USA \\
  \texttt{ttruong@openprotein.ai} \\
  \And
  Tristan Bepler \\
  OpenProtein.AI\\
  NY, USA \\
  \texttt{tbepler@openprotein.ai} \\
}

\begin{document}
\setcitestyle{numbers,square}
\renewcommand{\cite}[1]{\citep{#1}}

\maketitle

\begin{abstract}
Protein language models (PLMs) learn probability distributions over natural protein sequences. By learning from hundreds of millions of natural protein sequences, protein understanding and design capabilities emerge. Recent works have shown that scaling these models improves structure prediction, but does not seem to improve mutation understanding and representation quality for protein function prediction. We introduce PoET-2, a multimodal, retrieval-augmented protein foundation model that incorporates in-context learning of family-specific evolutionary constraints with optional structure conditioning to learn generative distributions over protein sequences. PoET-2 uses a hierarchical transformer encoder that is equivariant to sequence context ordering and a dual decoder architecture with both causal and masked language modeling objectives, allowing PoET-2 to operate in both fully generative and bidirectional representation learning modes. PoET-2 achieves state-of-the-art performance on zero-shot variant effect prediction, excelling at scoring variants with multiple mutations and challenging indel mutations. In supervised settings, PoET-2 embeddings outperform previous methods for learning sequence-function relationships, especially with small datasets. This work highlights the benefits of combining retrieval augmentation with multimodal, family-centric modeling for advancing protein foundation models. \footnote{PoET-2 code and model weights available at: https://github.com/OpenProteinAI/PoET-2}
\end{abstract}

\section{Introduction}
Proteins are polymer chains of amino acids that fold into 3-dimensional structures and carry out the vast majority of functions at the molecular level of life. Proteins catalyze chemical reactions, sense and respond to environmental signals, and defend against pathogens, among countless other functions. Their vast functional diversity arises from the astronomical space of possible amino acid sequences, which evolution has explored over billions of years through mutation and selection.

Accurate prediction of the effect of mutations on protein function is crucial for disease understanding, drug development, and protein engineering. Recent advances in protein language models (PLMs) have enabled more accurate zero-shot prediction of variant effects \cite{poet1, esm1v, trancepteve, tranception}. By learning to model probability distributions over natural protein sequences, PLMs output sequence likelihoods that capture relative fitness information and achieve remarkable correlation with functional and structural properties of proteins in deep mutational scanning and clinical mutation benchmarks \cite{proteingym}. However, several key challenges remain. 

\begin{itemize}

\item Most PLMs use masked language model-based approaches that are limited to prediction of single substitution mutations. These approaches are unable to predict the effect of insertions and deletions (indels), as well as epistatic effects in higher order mutations. 

\item PLMs are usually evaluated in the zero-shot setting, which is fundamentally limited to evaluating the correlation between estimates of sequence fitness and actual functional properties of proteins. However, in protein engineering campaigns, practitioners seek to learn directly from limited mutagenesis data to optimize for specific functional properties. PLMs evaluated in this supervised few-shot setting are promising \cite{bepler_overview, proteinnpt, esm1b}, but better data efficiency and generalization, particularly for sequence positions not observed during training, is still needed. 

\item Recent progress in PLMs have generally focused on scaling the number of parameters \cite{esm3}. However, increasing model capacity primarily seems to benefit only structure prediction \cite{esm2, esm3, xtrimopglm}, while showing neutral or even negative impacts on fitness modeling and function prediction \cite{poet1, misspec}. This raises concerns of loss of generalizability due to model memorization, while also making these models increasingly expensive to train and deploy for inference. Recent PLMs have explored alternative approaches that incorporate additional information via multi-modal approaches \cite{esm3, prosst, saprot}, or retrieval augmentation \cite{poet1}, but not both. 

\end{itemize}

To address these challenges, we propose PoET-2, a sequence-structure multimodal PLM that leverages retrieval-augmentation and dual training objectives to learn to generate new protein sequences conditioned on sequences and structures of homologs. PoET-2 combines three key ideas: 

\begin{itemize}
    
    \item \textbf{Multi-modality}: PoET-2 reasons over both sequence and structure. This enables conditioning on sequence and/or structural homologs, including structure-conditioned sequence generation from partially-observed backbone atomic structures
    \item \textbf{Retrieval-augmentation}: PoET-2 introduces a novel context-conditioning framework featuring a hierarchical attention architecture that is fully equivariant to the order of proteins in the context. This eliminates the need for multi-billion parameter models, while enabling in-context learning by allowing the model to be prompted with new sequences not present in the original training data. 
    \item \textbf{Dual training objectives}: PoET-2 is trained using both a causal language modeling objective for sequence generation and likelihood calculation, as well as a masked language modeling objective for bidirectional understanding and sequence embedding.
\end{itemize}

PoET-2's novel architecture achieves remarkable performance on downstream protein understanding and design tasks. It is capable of solving problems not possible with existing PLMs, including zero-shot indel and higher-order variant effect prediction, improving on previous methods by as much 20\%, on both deep mutational scanning and clinical datasets. Homology-augmented protein representations learned by PoET-2 also enable state-of-the-art accuracy in supervised few-shot function learning, reaching the performance of previous methods like Kermut \cite{kermut} with substantially less data and outperforming other PLMs by a large margin in the contiguous and modulo dataset splits. In ablation experiments, we find that structure conditioning primarily contributes to zero-shot prediction of stability while offering little to no benefit on tasks like clinical variant effect prediction and supervised function prediction. PoET-2 offers fast inference with an efficient footprint of only 182M parameters and minimal GPU requirements.

\section{Related Work}

\textbf{Zero-shot variant effect prediction} has advanced significantly by integrating information across sequence, structure, and evolution (homologs). Early progress was driven by single-sequence PLMs such as ESM \cite{esm1b,esm1v,esm2} and ProtT5 \cite{prottrans}, which, when trained at evolutionary scale, demonstrated strong correlations between sequence likelihoods and protein fitness. Concurrently, family-specific models emerged, focusing on narrower evolutionary contexts \cite{eve,deepsequence}. To capture broader evolutionary signals and enable knowledge transfer beyond single-family scopes, other models were trained across vast collections of distinct protein families. This was achieved through methods processing multiple sequence alignments (MSAs) e.g. MSA Transformer \cite{msa_transformer}, and later via models utilizing unaligned homologs e.g. PoET-1 \cite{poet1}. More recently, strategies for integrating structural information have been explored. These include using discrete structural tokens e.g. SaProt, ProSST, ESM3 \cite{saprot,prosst,esm3}, employing continuous geometric representations e.g. ProteinMPNN, ESM-IF \cite{gvp,proteinmpnn,esm_if}, and explicitly leveraging protein surface information e.g. S3F, RSALOR \cite{multiscale,rsalor}. Lastly, ensemble methods e.g. VenusREM \cite{venusrem} combine different methods to further enhance performance.

\textbf{Supervised variant effect prediction} commonly utilizes likelihoods or embeddings from PLMs as features for downstream models, enabling fitness prediction from limited experimental data \cite{hsu_likelihood,Yang2019_review,bepler_embeddings,bepler_overview,proteinnpt,kermut}. For instance, ProteinNPT \cite{proteinnpt} integrated MSA Transformer embeddings and zero-shot scores within a non-parametric Transformer architecture. Kermut \cite{kermut} further advanced this paradigm, achieving state-of-the-art results with a composite Gaussian Process (GP) kernel that incorporates features from multiple models, including ESM-2, ProteinMPNN, and AlphaFold2 \cite{alphafold2}.

\section{PoET-2}

\begin{figure}[htbp]
  \centering
   \includegraphics[width=0.85\linewidth]{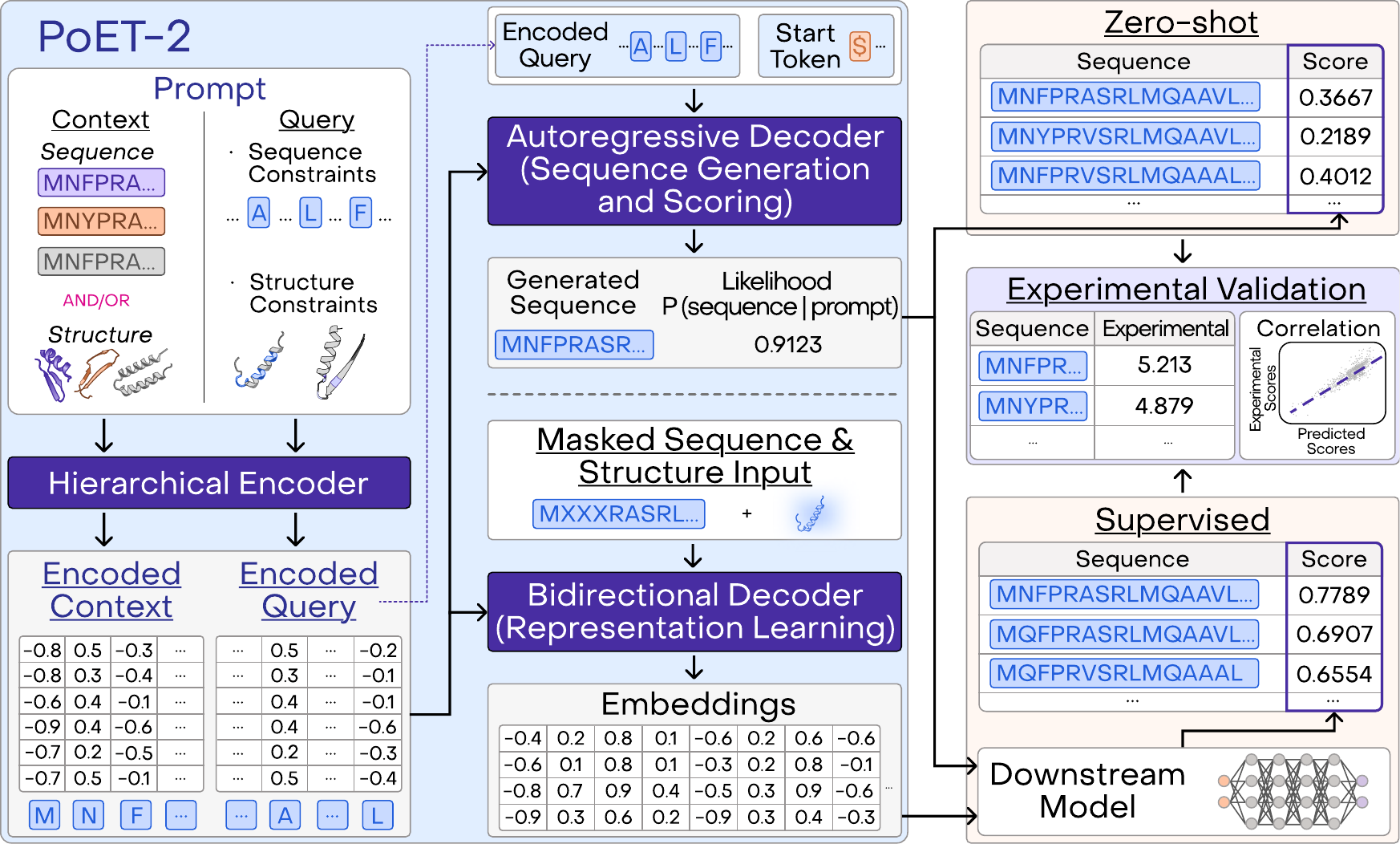}
  \caption{PoET-2 architecture and framework for zero-shot and supervised variant effect prediction. PoET-2 encodes a set of evolutionarily relevant proteins with an equivariant encoder, and decodes proteins with either of two decoders. Log-likelihoods from the autoregressive decoder are used for zero-shot prediction, and are combined with embeddings from the bidirectional decoder for supervised prediction.}
  \label{fig:poet_2}
\end{figure}

\subsection{Overview}
PoET-2 is a multimodal generative model of protein families, designed for controllable protein generation and representation learning. By jointly modeling the sequences and structures of proteins within a protein family, PoET-2 infers—through in-context learning—the underlying evolutionary constraints that give rise to the family's characteristic sequence features, structural architectures, and/or functional properties. These inferred evolutionary constraints, coupled with a flexible grammar for specifying explicit sequence and structural constraints, enable the controlled generation of novel family members, including variants with enhanced characteristics relevant to their function.

PoET-2 is implemented as an encoder-decoder Transformer \cite{transformer} with one encoder and two decoders (Figure \ref{fig:poet_2}). The encoder processes a user-provided \textbf{prompt} containing a \textit{set of proteins} that guides the two decoders toward generating novel proteins with desired characteristics. The prompt is processed in a fully protein order equivariant manner, and consists of two optional components:

\begin{itemize}
    \item \textbf{Context}: A protein family comprised of a set of proteins that the user believes are likely to exhibit at least one of the desired characteristics. 
    \item \textbf{Query}: A single, partially specified protein that specifies the sequence and/or structure at only a subset of residues; when used, the query constrains the model to generate only proteins containing those sequence or structural elements. Common uses of the query include specifying the protein length, the presence of signal peptides, the inclusion of active sites, or the structure of the entire protein backbone (i.e. inverse folding).
\end{itemize}

Together, the context and query provide a flexible grammar that allows sequence generation to be controlled both (1) \textit{implicitly}, via the context, which includes examples of proteins likely to exhibit desired characteristics, and (2) \textit{explicitly}, via the query, which specifies explicit sequence and structure constraints. Clever prompt engineering via careful selection of the context allows PoET-2 to focus on the evolutionary, structural, or functional constraints of only the subspace of relevant proteins.

The decoders, conditioning on the encoder's output, generate new proteins aligned with the prompted protein family. PoET-2 employs two distinct decoders for complementary strengths:

\begin{itemize}
    \item An \textbf{autoregressive decoder}, trained with a causal language modeling (CLM) objective, excels at generative tasks. By modeling the full joint probability distribution $P(\text{sequence}|\text{prompt})$, it allows for efficient, autoregressive generation of novel proteins and exact probabilistic scoring of sequence variants, including indels.
    \item A \textbf{bidirectional decoder}, trained with a masked language modeling (MLM) objective, specializes in representation learning. It produces powerful, context-aware embeddings where each residue's representation is informed by the entire sequence context (both preceding and succeeding residues). These rich representations capture deep structural and functional insights, crucial for tasks like structure and function prediction where understanding global dependencies is key.
\end{itemize}



Both decoders are also equivariant to the order of proteins in the prompt, meaning that PoET-2 as a whole is also equivariant. The complete loss function is thus composed of three components: $\mathcal{L}=\mathcal{L}_{\text{MLM encoder}} + \mathcal{L}_{\text{CLM decoder}} + \mathcal{L}_{\text{MLM decoder}}$.

The MLM encoder loss is the standard MLM loss applied to each protein in the prompt independently. The set of proteins is viewed as a sequence-of-sequences, where the order of proteins is arbitrary. Using the notation $x^{(i)}_j$ to denote the $j$th residue of $i$th protein in the prompt, the encoder loss is:

\begin{equation}
    \mathcal{L}_{\text{MLM encoder}} = -\mathbb{E}_{x,m_x}\left[\frac{1}{|m_x|}\sum_{i,j\in m_x}{\log p(x^{(i)}_j|x_{\backslash m_x})}\right]
\end{equation}

where $m_x$ is the set of masked positions in the sequence-of-sequences $x$.

The CLM decoder loss is the standard CLM loss additionally conditioned on (1) the prompt, $x_{\backslash m_x}$, and (2) an optional index, $q$, indicating the index of the sequence in the prompt to use as the query:

\begin{equation}
    \mathcal{L}_{\text{CLM decoder}} = -\mathbb{E}_{y,x,m_x,q}\left[\frac{1}{L_y}\sum_{i=1}^{L_y}{\log p(y_i|y_{<i},x_{\backslash m_x},q)}\right]
\end{equation}

Here, $y$ is a single sequence of length $L_y$; the notation $y_i$ refers to the $i$th residue of $y$.

Likewise, the MLM decoder loss is the standard MLM loss additionally conditioned on $x_{\backslash m_x}$ and $q$:

\begin{equation}
    \mathcal{L}_{\text{MLM decoder}} = -\mathbb{E}_{y,m_y,x,m_x,q}\left[\frac{1}{|m_y|}\sum_{i\in m_y}{\log p(y_i|y_{\backslash m_y},x_{\backslash m_x},q)}\right]
\end{equation}

\subsection{Architectural Details}

In this section, we introduce the architectural details of PoET-2 that are relevant to the capabilities demonstrated in this paper; see Appendix \ref{appendix:mic_architecture} for details relating to additional capabilities.

\subsubsection{Model Inputs}\label{model_inputs}

\textbf{Sequence} Protein sequences ($x_\text{seq}$) are tokenized with a single token per amino acid, a start token (\texttt{\$}) indicating the start of a sequence, a stop token (\texttt{*}) indicating the end of a sequence, a mask token (\texttt{X}) indicating a single residue with unknown identity, and a "gap" token (\texttt{-}) indicating zero or more residues with unknown identity.

\textbf{Structure} Protein structures backbones (N, C$\alpha$, and C atoms) are encoded in a roto-translation invariant way using two representations:

\begin{itemize}
    \item Pairwise C$\alpha$ distances ($D$): all pairwise C$\alpha$-C$\alpha$ distances, discretized into 128 bins: 125 equal-width bins (2.5Å-48Å), one bin for distances > 48Å, one for low-confidence pairs (pLDDT < 70), and one for missing or masked pairs.
    \item Local structure backbone distances ($x_\text{atomb}$): the 36 pairwise distances between the backbone atoms of the residue being encoded and the residues to its left and right along the sequence. These distances capture backbone information not fully recoverable from just $D$. 
\end{itemize}


\textbf{Predicted structure confidence (pLDDT)} As PoET-2 is trained only on predicted structures in AlphaFoldDB (AFDB) \cite{afdb,afdb-2}, it also takes as input the predicted structure confidence, pLDDT, at each residue ($x_{\text{plddt}}$).

\subsubsection{Input Embedding}
The encoder and decoders share a common input embedding space that fuses the sequence ($x_\text{seq}$), local structure backbone ($x_\text{atomb}$), and pLDDT ($x_\text{plddt})$ into a single continuous latent space via summation (Appendix Algorithm \ref{alg:embed_inputs}).

\subsubsection{Structure-based Attention Bias}\label{structure_based_bias}
To enhance structural information integration, PoET-2 employs a structure-based attention bias for all attention operations within individual protein sequences (but not between sequences) in both its encoder and decoders. This mechanism modifies attention scores by adding a learned bias term corresponding to the discretized pairwise C$\alpha$-C$\alpha$ distance bin ($D$) for each residue pair (Appendix Figure \ref{fig:structure_based_bias}). This approach is analogous to the relative positional bias in the T5 Transformer \cite{t5}; however, in PoET-2, the bias is determined by 3D structural proximity rather than linear sequence position.

\subsubsection{Encoder Architecture}\label{encoder_architecture}
PoET-2's encoder layers (Appendix Algorithm \ref{alg:encoder_layer}) adopt the architecture of standard Transformer encoder layers with modern tweaks (rotary positional encodings, SwiGLU over MLP, and RMSNorm over LayerNorm) \cite{transformer,rope,glu,rmsnorm}, and modifications to (1) ensure protein order equivariance, and (2) improve handling of structural inputs. To process sets of input proteins in an order equivariant way, it replaces the standard one stage attention mechanism with the two stage, hierarchical attention mechanism introduced by PoET-1 \cite{poet1}. Summarizing briefly,

\begin{itemize}
    \item In the first stage, attention is applied only between residues of individual input proteins. Structure-based attention bias (\S\ref{structure_based_bias}) is used in this stage.
    \item In the second stage, attention is applied between residues of all input proteins. Additionally, relative positional encodings between residues reflect the absolute positions within each protein, rather than the absolute position in the sequence-of-sequences.
\end{itemize}

Attention in PoET-2's encoder is fully bidirectional, enabling the entire encoder to be protein order equivariant. This is in contrast to PoET-1, whose decoder-only, autoregressive architecture permits equivariance only in each individual decoder layer, and not the entire decoder.

\subsubsection{Decoder Architecture}\label{decoder_architecture}
Similarly, PoET-2's decoder layers (Appendix Algorithm \ref{alg:decoder_layer}) adopt the architecture of the standard Transformer decoder layer with modifications. The modifications to the attention operations are as follows:

\begin{itemize}
    \item In the first, self-attention stage, structure-based attention bias (\S\ref{structure_based_bias}) is used.
    \item In the second, cross-attention stage, protein order equivariance is maintained by using the same relative position scheme as in the second attention stage of the encoder.
\end{itemize}

Additionally, when the prompt includes a query, the input embedding of the decoder is modified to encode which protein in the prompt should be used as the query. The modified input embedding of each residue is simply the average of the unmodified input embedding, and the embedding, produced by the encoder, of corresponding residue of the query in the prompt.

\subsection{Hyperparameters}
PoET-2 is 182 million parameter model, structured with 12 layers and a 1024 hidden dimension in its encoder and decoders. Weights are tied between these modules to enhance parameter efficiency and promote shared representation learning. This approach leverages the ideal representational capabilities of bidirectional modeling, while also benefiting from the strong representations achievable with autoregressive architectures \cite{proteinnpt,poet-supervised-blog-post}.

\subsection{Training Data}
PoET-2 is trained on 62 million sets of homologous sequences. Each set corresponds to a sequence in UniRef50 Version 2304 \cite{uniprot}, and contains all of its homologs in UniRef50 found using Diamond \cite{diamond}. Each sequence may optionally be associated with a predicted structure from AFDB by matching on the UniRef100 identifier. To ensure that PoET-2 sees a variety of prompts during training, and to reduce the risk of overfitting, sequence and structure are randomly masked (Appendix \ref{appendix:poet_2_training_details}).

\section{Experiments}\label{experiments}
Variant effect prediction is the task of predicting the effect of mutations on the ability of a protein to perform its function. To evaluate PoET-2's zero-shot and supervised variant effect prediction capabilities, we utilize the ProteinGym benchmark \cite{proteingym}. ProteinGym assesses the performance of a model by measuring its ability to predict variant effect in two types of datasets: (1) deep mutational scanning (DMS) datasets, which encompass over 200 distinct assays measuring the effect of mutations on a wide variety of proteins and protein functions spanning the tree of life, and (2) clinical datasets measuring the pathogenicity of mutations on >2,500 human genes.

Following ProteinGym conventions, we use Spearman's rank correlation coefficient ($\rho$) between experimental measurements and predicted fitness as the primary metric for continuous variables, and area under the receiver operating curve (AUROC) for binary variables.

As baselines, we use a subset of the evaluated methods recorded in ProteinGym as of May 12 2025, including the top methods for each benchmark subset, and methods covering a wide variety of approaches e.g. using structure information or not, using sequence homologs or not, etc.

\subsection{PoET-2 achieves state-of-the-art zero-shot performance for challenging mutations and clinical variants}\label{zero_shot_methods}

In zero-shot variant effect prediction, models must predict the effect mutations on protein fitness without training on experimental data for the specific protein or function of interest. To score the fitness of a mutated variant sequence relative to its wild-type (WT) using PoET-2, we use the log likelihood ratio (LLR) under PoET-2's CLM decoder: $\log{P(\text{variant}|\text{prompt})}-\log{P(\text{WT}|\text{prompt})}$. To optimize predictions, we employ several prompt engineering and scoring adjustment strategies:

\begin{itemize}
\item \textbf{Ensembling Prompts}: Following PoET-1 \cite{poet1}, we average LLRs obtained from multiple prompts. Each prompt contains a different subsample of WT homologs identified by searching UniRef100 using the ColabFold MSA protocol \cite{colabfold}. Sampling parameters are detailed in Appendix \ref{appendix:prompt_engineering_zero_shot}.
\item \textbf{Structure Conditioning}: We employ two ways of incorporating structure in the prompt. The utility of structural information varies by task, as discussed in \S\ref{structure_role}. 
\item \textbf{Length Adjustment for Indels}: To correct for potential biases in autoregressive models towards shorter sequences, we score indel variants using a length-adjusted log likelihood ratio (Appendix \ref{appendix:length_adjusted_llr}).
\end{itemize}

\begin{table}[htbp]
\caption{Performance (Spearman $\rho$) on zero-shot DMS substitutions and indels benchmarks. N/A indicates that the model cannot score indels.}
\label{tab:dms_unsupervised}
\centering
\resizebox{\textwidth}{!}{
\begin{tabular}{lrccccccc}
\toprule
\multirow{2}{*}{\textbf{Model}} & \multirow{2}{*}{\textbf{\# Param}} & \multicolumn{2}{c}{\textbf{Model Inputs}} & \multicolumn{4}{c}{\textbf{Substitutions By MSA Depth}} & \multirow{2}{*}{\textbf{Indels}} \\
\cmidrule(lr){3-4} \cmidrule(lr){5-8}
& & \textbf{Struct.} & \textbf{MSA} & \textbf{Low} & \textbf{Medium} & \textbf{High} & \textbf{All} & \\
\midrule
ESM-2 & 650M &  &  & 0.340 & 0.410 & 0.513 & 0.415 & N/A \\
ESM C & 300M & $\times$ & $\times$ & 0.338 & 0.401 & 0.519 & 0.407 & N/A \\
ProGen2 M & 764M &  &  & 0.305 & 0.390 & 0.422 & 0.379 & 0.466 \\
ProGen2 XL & 6.4B &  &  & 0.322 & 0.411 & 0.442 & 0.390 & 0.430 \\
\midrule
SaProt & 650M &  &  & 0.397 & 0.446 & 0.546 & 0.457 & N/A \\
ESM-3 Open & 1.4B & $\checkmark$ & $\times$ & 0.402 & 0.465 & 0.575 & 0.467 & N/A \\
ProSST & 110M &  &  & 0.468 & 0.506 & 0.581 & 0.508 & N/A \\
\midrule
MSA Transformer & 100M &  &  & 0.375 & 0.456 & 0.480 & 0.431 & N/A \\
TranceptEVE L & 700M & $\times$ & $\checkmark$ & 0.434 & 0.473 & 0.491 & 0.456 & 0.414 \\
GEMME & N/A &  &  & 0.445 & 0.474 & 0.494 & 0.455 & N/A \\
PoET-1 & 200M &  &  & 0.479 & 0.477 & 0.511 & 0.470 & 0.517 \\
\midrule
S3F-MSA & 910M &  &  & 0.470 & 0.509 & 0.547 & 0.496 & N/A \\
VenusREM & 110M & $\checkmark$ & $\checkmark$ & 0.498 & 0.524 & 0.578 & 0.519 & N/A \\
PoET-2 & 182M &  &  & 0.488 & 0.507 & 0.555 & 0.500 & \textbf{0.566} \\
\cdashline{1-9}
PoET-2 + VenusREM & 292M &  &  & \textbf{0.528} & \textbf{0.550} & \textbf{0.593} & \textbf{0.543} & N/A \\
\bottomrule
\end{tabular}
} 
\end{table}

\subsubsection{PoET-2 significantly advances our ability to predict the effects of indels and higher-order mutations}
Predicting the effects of complex mutations such as insertions/deletions (indels) and higher-order substitutions is a challenge for many PLMs. State-of-the-art structure-aware predictors based on masked language modeling (e.g. VenusREM \cite{venusrem}, S3F-MSA \cite{multiscale}) operate on fixed-length sequences, which prevents them from directly scoring length-altering indels. Furthermore, for higher-order mutations, they assume additive effects across mutated positions and therefore cannot fully model epistatic interactions. In contrast, PoET-2's autoregressive decoder conditions on both sequence and structure to model the full joint probability of a sequence, $P(\text{sequence}|\text{prompt})$. This approach naturally handles variable sequence lengths and epistatic effects.

\textbf{DMS indels} On the ProteinGym DMS indels benchmark (Indels column, Table \ref{tab:dms_unsupervised}), PoET-2 significantly outperforms all existing models. It achieves a substantial improvement of $\Delta\rho\approx 0.05$ ($p<1e-5$) over PoET-1, the previous best. Compared to the top-performing non-PoET model, PoET-2 demonstrates an even larger lead of $\Delta\rho\approx 0.10$ ($p<1e-5$), an improvement of over 20\%.

\textbf{DMS higher-order substitutions} For higher-order substitution mutations (Table \ref{tab:mutation_depth}), PoET-2 demonstrates exceptional performance, particularly for variants with three or more mutations. When compared to VenusREM, the state-of-the-art model on the overall DMS substitutions benchmark, PoET-2 achieves substantial gains on these more complex variants ($\Delta\rho\approx 0.09$ for 3 mutations, $\Delta\rho\approx 0.10$ for 4 mutations, and $\Delta\rho\approx 0.075$ for 5+ mutations).

\begin{table}[htbp]
\resizebox{0.95\textwidth}{!}{
\begin{minipage}{0.64\linewidth}
\caption{Performance (Spearman $\rho$) on zero-shot DMS substitutions benchmark, by mutation depth (i.e. by number of substitutions).}
\label{tab:mutation_depth}
\centering
\begin{tabular}{lccccc}
\toprule
& \multicolumn{5}{c}{\textbf{Substitutions by Mutation Depth}} \\
\cmidrule{2-6}
\textbf{Model} & \textbf{1} & \textbf{2} & \textbf{3} & \textbf{4} & \textbf{5+} \\
\midrule
ESM-2 & 0.422 & 0.245 & 0.203 & 0.160 & 0.220 \\
ESM C & 0.417 & 0.255 & 0.189 & 0.150 & 0.217 \\
ProGen2 M & 0.372 & 0.126 & 0.149 & 0.131 & 0.178 \\
ProGen2 XL & 0.370 & 0.138 & 0.219 & 0.200 & 0.261 \\
\midrule
SaProt & 0.460 & 0.310 & 0.271 & 0.268 & 0.337 \\
ESM-3 Open & 0.493 & 0.335 & 0.303 & 0.284 & 0.365 \\
ProSST & 0.523 & 0.391 & 0.316 & 0.274 & 0.334 \\
\midrule
MSA Transformer & 0.427 & 0.216 & 0.358 & 0.365 & 0.401 \\
TranceptEVE L & 0.446 & 0.274 & 0.349 & 0.327 & 0.385 \\
GEMME & 0.449 & 0.273 & 0.329 & 0.338 & 0.419 \\
PoET-1 & 0.467 & 0.295 & 0.412 & 0.393 & 0.421 \\
\midrule
S3F-MSA & 0.501 & 0.330 & 0.377 & 0.343 & 0.387 \\
VenusREM & 0.536 & 0.394 & 0.352 & 0.320 & 0.372 \\
PoET-2 & 0.508 & 0.355 & \textbf{0.444} & \textbf{0.419} & \textbf{0.447} \\
\cdashline{1-6}
\makecell[l]{PoET-2\\+ VenusREM} & \textbf{0.558} & \textbf{0.400} & 0.442 & 0.411 & 0.441 \\
\bottomrule
\end{tabular}
\end{minipage}
\hspace{5mm}
\begin{minipage}{0.36\linewidth}
\caption{Performance (AUROC) on zero-shot clinical substitutions and indels benchmarks. N/A indicates not applicable, whereas a dash (--) indicates applicable, but not computed.}
\label{tab:clinical}
\centering
\begin{tabular}{lcc}
\toprule
\textbf{Model} & \textbf{Subs.} & \textbf{Indels} \\
\midrule
Progen2 M & -- & 0.846 \\
Progen2 L & -- & 0.851 \\
Progen2 XL & -- & 0.842 \\
\midrule
RITA M & -- & 0.892 \\
RITA L & -- & 0.922 \\
RITA XL & -- & 0.916 \\
\midrule
PROVEAN & 0.886 & 0.927 \\
\midrule
ESM-1b & 0.892 & N/A \\
\midrule
TranceptEVE L & 0.920 & 0.857 \\
\midrule
PoET-1 & 0.920 & 0.934 \\
PoET-2 & \textbf{0.928} & \textbf{0.952} \\
\bottomrule
\end{tabular}
\end{minipage}
} 
\end{table}

\subsubsection{PoET-2 complements existing methods for substitution mutations}

\textbf{DMS substitutions} On the DMS substitutions benchmark (Table \ref{tab:dms_unsupervised}), PoET-2 achieves performance comparable to VenusREM \cite{venusrem}, the current state-of-the-art. VenusREM is an ensemble model combining ProSST \cite{prosst}, a structure-aware protein language model (PLM), with a Position-Specific Scoring Matrix (PSSM) derived from evolutionary alignments \cite{potts}. While PoET-2 slightly trails VenusREM on the primary Spearman correlation metric ($p<1e-3$), it demonstrates  superior or comparable performance on metrics emphasizing the prediction of beneficial mutations, such as normalized discounted cumulative gain (NDCG, 0.786 vs 0.766 for VenusREM, $p<1e-5$; Appendix \ref{appendix:zero_shot_detailed_results}). NDCG scores whether a model gives its highest scores to the sequences with highest fitness, indicating that PoET-2 is slightly better at identifying the most efficacious mutations versus ranking middling and deleterious ones as accurately.

A simple ensemble combining PoET-2 and VenusREM (Appendix \ref{appendix:zero_shot_ensembles}) consistently outperforms both individual models and all other existing methods across all metrics ($p<1e-5$; Table \ref{tab:dms_unsupervised}, Appendix \ref{appendix:zero_shot_detailed_results}). This suggests PoET-2 and VenusREM capture distinct, complementary fitness signals. The ensemble shows robust performance across diverse protein subgroups, including those with varying MSA depths (Table \ref{tab:dms_unsupervised}) and assay functions (Appendix \ref{appendix:zero_shot_detailed_results}). However, on higher-order substitutions (3 or more mutations), PoET-2 alone surpasses the ensemble's performance (Table \ref{tab:mutation_depth}), underscoring its strong intrinsic capability to model these complex mutational effects.

\subsubsection{PoET-2 improves prediction of clinical variant pathogenicity}

PoET-2 significantly improves our ability to distinguish between pathogenic and benign human protein mutations on the ProteinGym clinical benchmark (Table \ref{tab:clinical}). Compared to PoET-1, the next best model, PoET-2 improves AUROC by $0.008$ on the substitutions benchmark ($p<9e-5$) and by a substantial $0.018$ on the indels benchmark ($p<3e-5$), establishing a new state-of-the-art for both.

\subsection{PoET-2 embeddings and likelihoods enhance supervised learning of sequence-function relationships}
While zero-shot prediction is valuable, protein engineering often involves learning from limited experimental data. We evaluate PoET-2's utility in this supervised setting on ProteinGym's primary supervised DMS benchmark, which focuses on single-site substitutions across all 217 DMS assays. This benchmark assesses generalization ability across three cross-validation (CV) schemes, varying in difficulty based on the relationship between training and test set mutation locations. In the random fold, mutations are distributed randomly across five CV folds. In the modulo fold, protein positions are assigned to one of five CV folds using a modulo-based strategy i.e. every fifth position belongs to the same fold. In the contiguous fold, the protein sequence is divided into five contiguous, equal-length segments, each constituting a CV fold.


We employ a Gaussian Process (GP) regression model to predict fitness. The GP uses a product kernel combining two Matérn 5/2 kernels: one operating on protein embeddings from PoET-2's MLM decoder, and the other on LLRs from the CLM decoder (as used in zero-shot prediction). This kernel was chosen for its relative simplicity and minimal hyperparameter tuning requirements, but it may not be optimal for all scenarios, such as predicting the effects of multi-mutation variants (Appendix \ref{appendix:supervised_gp_kernels}). Predictions are ensembled across GPs trained on features from different sequence-only prompts; structure was omitted from prompts because they did not improve supervised results (\S\ref{structure_role}).


\subsubsection{PoET-2 improves supervised variant effect prediction across diverse generalization regimes}

Our PoET-2 based GP model (PoET-2 GP) substantially outperforms the previous state-of-the-art, Kermut \cite{kermut}, in all cross-validation folds on both Spearman correlation and mean squared error metrics (Table \ref{tab:dms_supervised}, $p<1e-5$). For example, PoET-2 GP achieves an average Spearman $\rho$ of 0.693, compared to 0.664 for Kermut. This improved performance is consistent across various protein and assay subgroups (Appendix \ref{appendix:supervised_detailed_results}).

\begin{table}[htbp]
\caption{Performance on supervised DMS substitutions benchmark.}
\label{tab:dms_supervised}
\centering
\begin{tabular}{lcccccccc}
\toprule
 & \multicolumn{4}{c}{\textbf{Spearman $\rho$ ($\uparrow$)}} & \multicolumn{4}{c}{\textbf{Mean square error ($\downarrow$)}} \\
\cmidrule(lr){2-5} \cmidrule(lr){6-9}
\textbf{Model} & \textbf{Rand.} & \textbf{Mod.} & \textbf{Contig.} & \textbf{Avg.} & \textbf{Rand.} & \textbf{Mod.} & \textbf{Contig.} & \textbf{Avg.} \\
\midrule
ProteinNPT       & 0.741          & 0.588          & 0.529          & 0.619          & 0.441          & 0.765          & 0.856          & 0.687 \\
\midrule
Kermut           & 0.746          & 0.635          & 0.613          & 0.664          & 0.413          & 0.649          & 0.697          & 0.586 \\
\midrule
ESM-2 (650 M) GP & 0.749          & 0.573          & 0.549          & 0.624          & 0.404          & 0.720          & 0.768          & 0.630 \\
ESM C GP         & 0.747          & 0.605          & 0.573          & 0.642          & 0.398          & 0.660          & 0.716          & 0.592 \\
PoET-2 GP        & \textbf{0.773} & \textbf{0.661} & \textbf{0.645} & \textbf{0.693} & \textbf{0.370} & \textbf{0.602} & \textbf{0.647} & \textbf{0.540} \\
\bottomrule
\end{tabular}
\end{table}

\subsubsection{PoET-2 has exceptional data efficiency for few-shot function learning}
A critical aspect of practical protein engineering is a model's ability to learn effectively from limited experimental data. To assess data efficiency and compare the utility of different foundation models, we benchmark GP models using identical kernel functions but features derived from different foundation models, including PoET-2, PoET-1 \cite{poet1}, ESM-2 \cite{esm2}, and ESM C \cite{esmc}. To simulate smaller training set sizes, we systematically vary the training data for each assay by subsampling its available training points, targeting specific sizes from as few as 10 points up to the maximum available.

\begin{figure}[htbp]
\centering
\includegraphics[trim={0 0 0 0},clip,width=\linewidth]{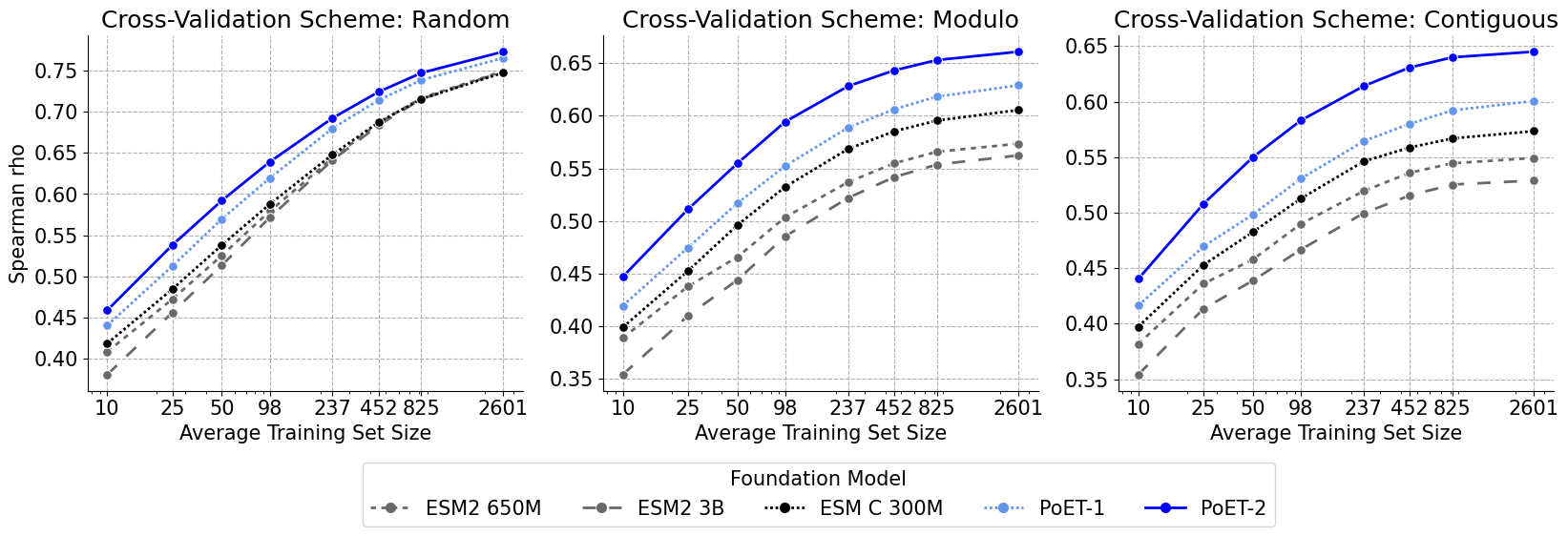}
\vspace*{-5mm}
\caption{Impact of training set size on the performance of Gaussian Process (GP) models leveraging various foundation models, evaluated on the supervised DMS substitutions benchmark.}
\label{fig:dms_supervised_data_titration}
\end{figure}

PoET-2 GP consistently outperforms GPs based on other foundation models across all evaluated training set sizes and cross-validation schemes (Figure \ref{fig:dms_supervised_data_titration}). PoET-2's advantage is most pronounced in the challenging contiguous cross-validation split, where PoET-2 GP trained with at most 100 data points matches the performance of ESM C GP (the strongest non-PoET based GP) trained with the maximum training set size (averaging \textasciitilde 2600 data points across all assays). Moreover, PoET-2 GP trained with at most 250 data points achieves performance comparable to Kermut trained with the maximum training set size, demonstrating exceptional data efficiency for practical protein engineering applications compared to the existing state-of-the-art.

\subsection{Structure conditioning improves zero-shot prediction, but has limited impact on supervised prediction}\label{structure_role}

To leverage PoET-2's multimodality, we explore two methods for incorporating structure in the prompt. First, we include the predicted structures of homologous proteins in the context. Second, following prior work showing that inverse-folding likelihood correlates with fitness \cite{proteingym}, we use a query consisting of only the WT structure to score a variant's likelihood of adopting the same fold.


In the zero-shot setting, both strategies improve performance on DMS substitutions over using sequence alone (Table \ref{tab:unsupervised_dms_prompt_engineering}). Including predicted structures in the context and using the inverse-folding query each individually improve performance, with the best results coming from ensembling these different strategies. Consistent with prior work, this highlights that structural priors are highly informative for zero-shot prediction. This is particularly evident for stability-related assays, where PoET-2 achieves its largest performance gains over its sequence-only predecessor, PoET-1 (Table \ref{tab:dms_unsupervised}). For other tasks like clinical variant prediction, however, the benefit is less clear, with the inverse-folding query being slightly detrimental (Appendix Table \ref{tab:unsupervised_clinical_prompt_engineering}). A detailed analysis of these strategies on both DMS and clinical benchmarks is in Appendix \ref{appendix:prompt_engineering_zero_shot}.

\begin{table}[htbp]
\centering
\caption{Performance (Spearman's $\rho$) of different strategies for including structure in the prompt on the zero-shot DMS benchmarks.}
\label{tab:unsupervised_dms_prompt_engineering}
\begin{tabular}{c|cccc}
\toprule
\multirow{2}{*}{\textbf{Strategy}} & \multicolumn{2}{c}{\textbf{Prompt}} & \multirow{2}{*}{\textbf{Substitutions}} & \multirow{2}{*}{\textbf{Indels}} \\
\cmidrule{2-3}
& \textbf{Context Modalities} & \textbf{Query} & \\
\midrule
A & Sequence & None & 0.47534 & 0.55589 \\
B & Sequence and Structure & None & 0.48374 & \textbf{0.56666} \\
C & Sequence & Structure of WT & \textbf{0.49128} & N/A \\
D & Sequence and Structure & Structure of WT & 0.48927 & N/A \\
\midrule
E & \multicolumn{2}{c}{\makecell[c]{Ensemble of A and B \\ {\small(Different contexts, No query)}}} & 0.48260 & \textbf{0.56606} \\
\addlinespace[1.5pt]
\cdashline{1-5}
\addlinespace[1.5pt]
F & \multicolumn{2}{c}{\makecell[c]{Ensemble of C and D \\ {\small(Different contexts, With query)}}} & 0.49256 & N/A \\
\addlinespace[1.5pt]
\cdashline{1-5}
\addlinespace[1.5pt]
G & \multicolumn{2}{c}{\makecell[c]{Ensemble of A and C \\ {\small(Sequence only context, Different queries)}}} & 0.49632 & N/A \\
\addlinespace[1.5pt]
\cdashline{1-5}
\addlinespace[1.5pt]
H & \multicolumn{2}{c}{\makecell[c]{Ensemble of B and D \\ {\small(Sequence and structure context, Different queries)}}} & 0.49987 & N/A \\
\addlinespace[1.5pt]
\cdashline{1-5}
\addlinespace[1.5pt]
I & \multicolumn{2}{c}{Ensemble of A, B, C, and D} & 
\textbf{0.49989} & N/A \\
\bottomrule
\end{tabular}%
\end{table}

In contrast, for supervised learning, including structural information offers little to no benefit (Appendix \ref{appendix:prompt_engineering_supervised}). This suggests that PoET-2's embeddings already implicitly encode critical structural information, and as a result, our current supervised model does not gain additional predictive power from explicit structure conditioning. Unlocking further improvements may require more sophisticated methods capable of leveraging this explicit structural data to refine an already strong baseline. 

\section{Conclusion and Limitations}\label{conclusion_and_limitations}

PoET-2 is a multimodal, retrieval-augmented protein language model. PoET-2 can learn from unaligned sequences and structure in-context and directly condition on atomic backbone structure for protein sequence generation and representation learning. As a result, PoET-2 achieves state-of-the-art performance for zero-shot indel and higher order mutation effect prediction, clinical variant effect prediction, and supervised variant effect prediction. However, it lags slightly behind recent structure-based masked language models on single mutant effects on DMS datasets. This seems to be largely driven by these models' superior performance on stability datasets. However,  ensembling PoET-2 with VenusREM produces a predictor that outperforms all previous models on ProteinGym's DMS benchmark, suggesting there is still orthogonal information being learned by these methods. Structure-based methods have increasingly been adopting discrete structure tokenizations whereas PoET-2 operates directly on backbone atoms. We also find that structure conditioning is only helpful for some problems, in particular stability prediction in the ProteinGym DMS datasets. For clinical variant effect prediction and supervised learning, structure conditioning offers little to no benefit. In principle, predicted structure information should already be encoded in protein language model-based representations. PoET-2 offers state-of-the-art performance in a compact 182M parameter footprint. We expect PoET-2 to become a core part of protein machine learning and engineering workflows.

\newpage
\section*{Acknowledgments}
We thank Grace Yeo for her valuable assistance in revising and editing the manuscript.

\bibliography{references}

\newpage
\appendix
\section{Broader Impact}\label{appendix:broader_impacts}

PoET-2, a multimodal, retrieval-augmented protein model, aims to improve the prediction of mutational effects and enable controllable protein design by learning from sequence, structure, and evolutionary family context. This work can accelerate beneficial applications such as designing novel enzymes, therapeutics, and more stable proteins. PoET-2's enhanced data efficiency in supervised learning may also broaden access to advanced protein engineering, especially in data-limited settings. While PoET-2 is a foundational research tool, advanced capabilities in understanding and designing proteins could theoretically be misused, for example, in the development of dangerous drugs. We expect that future work will serve to address and mitigate such concerns.

\section{PoET-2 Architecture}\label{appendix:poet_2_architecture}

This section elaborates on PoET-2's architecture, supplementing the description in the main text.

\paragraph{Notation} To refer to a specific residue in a sequence-of-sequences $x$, we use the same notation as in the main text i.e. we use $x^{(i)}_j$ to denote the $j$th residue of the $i$th sequence in $x$. In general, the superscript ${}^{(i)}$ is used to refer to the $i$th sequence. For example, in the main text, we use $D$ to refer to the matrix of pairwise discretized C$\alpha$ distances of a protein. Thus, in the context of a sequence-of-sequences, the notation $D^{(i)}$ refers to the pairwise discretized C$\alpha$ distances of the $i$th sequence in the sequence-of-sequences.

\subsection{Input Embedding}

The encoder and decoders share a common input embedding space. This space fuses representations derived from the input sequence ($x_\text{seq}$), the local structure backbone coordinates ($x_\text{atomb}$), and the pLDDT scores ($x_\text{plddt}$). Both $x_\text{atomb}$ and $x_\text{plddt}$ can contain entries that are masked, for instance, due to missing structural information, padding, or masking specified by a user. For each of these features, a corresponding binary mask (e.g., $x_\text{atomb\_mask}$, $x_\text{plddt\_mask}$) is provided, where a value of 1 indicates an observed or valid entry, and 0 indicates a masked or invalid entry. To process these potentially masked inputs, the algorithm first applies the respective binary mask to the feature data by setting the values at masked positions to zero. Subsequently, the binary mask itself is concatenated as an additional feature channel to this modified data. These augmented representations for $x_\text{atomb}$ and $x_\text{plddt}$ are then linearly projected into the target embedding dimension. The sequence $x_\text{seq}$ is embedded directly. Finally, these three resulting embeddings are summed to form the single continuous latent representation (Algorithm \ref{alg:embed_inputs}).

\begin{algorithm}[H]
\caption{\texttt{embed\_inputs} -- embeds a single sequence or a sequence-of-sequences}\label{alg:embed_inputs}
\begin{algorithmic}[1]
\Require $x_\text{inputs}=\{x_\text{seq}\in\{1..28\}^{L_x}, x_\text{plddt}\in[0,100]^{L_x},x_\text{plddt\_mask}\in\{0,1\}^{L_x}, x_\text{atomb}\in\mathbb{R}^{L_x\times36}, x_\text{atomb\_mask}\in\{0,1\}^{L_x\times36}$
\vspace{1mm}
\LeftComment Handle masks for $x_\text{plddt}$ and $x_\text{atomb}$ by applying the masks and concatenating along feature dimension
\State $z_{\text{plddt}}$ = \text{Concat}(($x_\text{plddt}*x_\text{plddt\_mask}$, $x_\text{plddt\_mask}$), \text{dim=1})\Comment{$[0,100]^{L_x\times2}$}
\State $z_{\text{atomb}}$ = \text{Concat}(($x_\text{atomb}*x_\text{atomb\_mask}$, $x_\text{atomb\_mask}$), \text{dim=1})\Comment{$\mathbb{R}^{L_x\times72}$}
\LeftComment Embed inputs individually
\State $z_{\text{seq}} = \text{Embed}(x_{\text{seq}})$ \Comment{$\mathbb{R}^{L_x\times d}$}
\State $z_{\text{plddt}} = \text{Linear}(z_{\text{plddt}})$ \Comment{$\mathbb{R}^{L_x\times d}$}
\State $z_{\text{atomb}} = \text{Linear}(z_{\text{atomb}})$ \Comment{$\mathbb{R}^{L_x\times d}$}
\State $z_{\text{seqid}} = \text{Linear}(z_{\text{seqid}})$ \Comment{$\mathbb{R}^{L_x\times d}$}
\State \textbf{return} $z_{\text{seq}} + z_{\text{plddt}} + z_{\text{atomb}} + z_{\text{seqid}}$
\end{algorithmic}
\end{algorithm}

\subsection{Structure-based Attention Bias}

\begin{figure}[H]
\centering
\includegraphics[width=\linewidth]{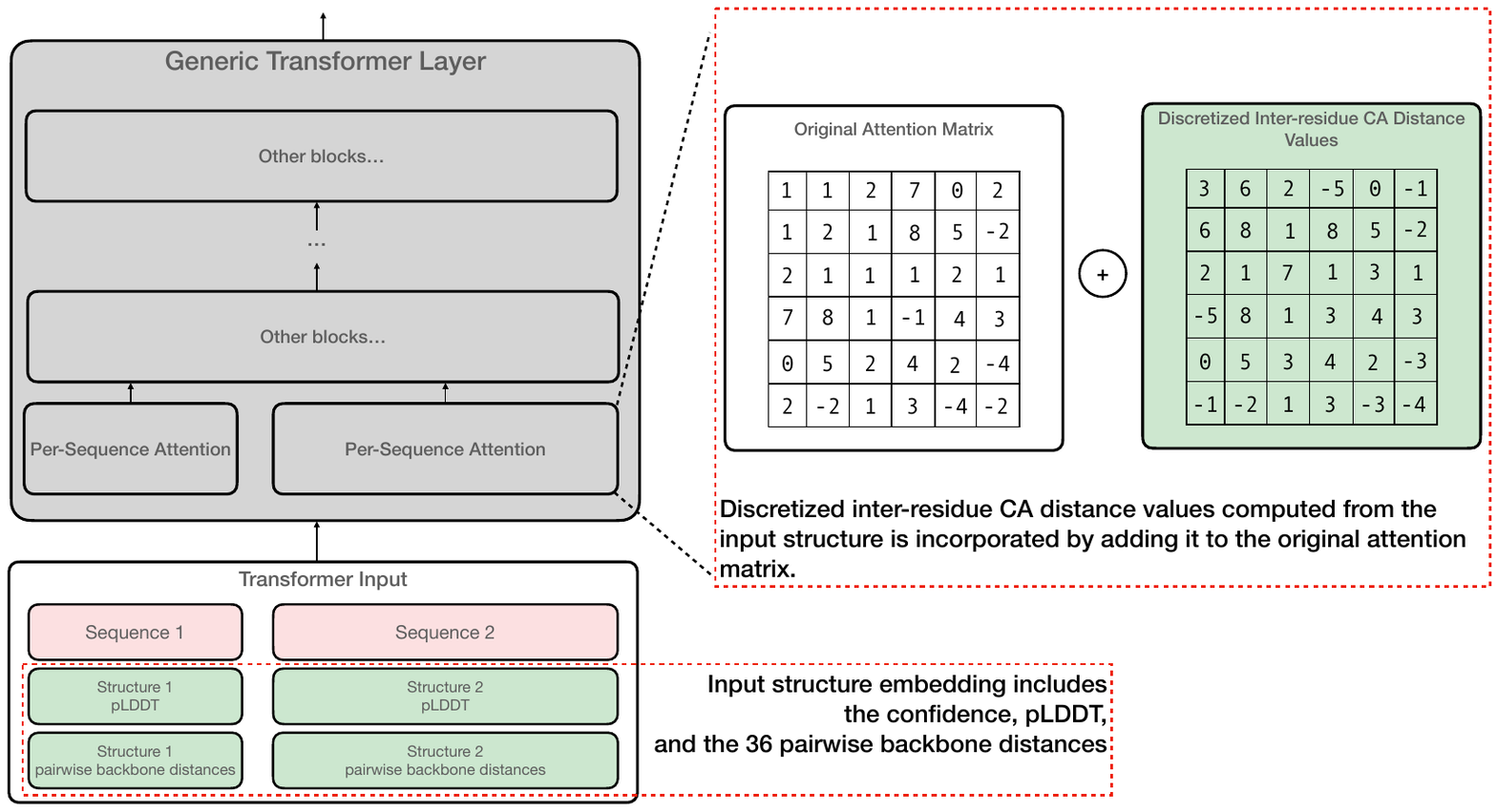}
\caption{Structure-based attention bias}
\label{fig:structure_based_bias}
\end{figure}

See the main text (\S\ref{structure_based_bias}) for a description of the structure-based attention bias. Figure \ref{fig:structure_based_bias} visualizes the application of the structure-based attention bias.

\subsection{Encoder Architecture}

After transforming the raw inputs $x_\text{inputs}$ into embeddings, the encoder (Algorithm \ref{alg:encoder}) further transforms the embeddings by applying $n_\text{layers}$ of protein order equivariant encoder layers (Algorithm \ref{alg:encoder_layer}). The encoder has two outputs: per residue embeddings of the prompt, $h_\text{encoder}$, that are used in the decoders, and per residue sequence logits, $z_\text{seq}$, that are used to compute the encoder MLM loss $\mathcal{L}_\text{MLM encoder}$.

\begin{algorithm}
\caption{\texttt{encoder\_layer} (for brevity, this algorithm describes only a single attention head, but can be extended to multiple attention heads in the normal fashion)}\label{alg:encoder_layer}
\begin{algorithmic}[1]
\Require $\forall i\in\{1..\text{N}\}, j\in\{1..L_{x_i}\}$
\item[\hspace{1em}\textbullet~prompt embedding $h^{(i)}_{j}\in\mathbb{R}^d$]
\item[\hspace{1em}\textbullet~pairwise discretized C$\alpha$ distances $D^{(i)}_{m,n}\in\{1..128\}, m\in\{1..L_{x_i}\}, n\in\{1..L_{x_i}\}$]
\item[]
\LeftComment First, apply self-attention with structure-based attention bias to each sequence individually
\State $f=\text{RMSNorm}(h)$ \Comment{$\mathbb{R}^{L_x\times d}$}
\State $q^{(i)}_j=\text{RoPE}(\text{Linear}(f^{(i)}_j), j) \hspace{1mm} \forall i,j$ \Comment{$\mathbb{R}^d$}
\State $k^{(i)}_j=\text{RoPE}(\text{Linear}(f^{(i)}_j), j) \hspace{1mm} \forall i,j$ \Comment{$\mathbb{R}^d$}
\State $v=\text{Linear}(f)$\Comment{$\mathbb{R}^{L_x\times d}$}
\LeftComment Compute attention score with structure-based bias
\State $A^{(i)}_{m,n}={q^{(i)}_m}^Tk^{(i)}_n+\texttt{structure\_bias}(D^{(i)}_{m,n})$\item[]\Comment{$\mathbb{R}$}
\State $f^{(i)}=f^{(i)}+\text{softmax}(\frac{A^{(i)}}{\sqrt{d}})v^{(i)} \hspace{1mm} \forall i$ \Comment{$\mathbb{R}^{L_{x^{(i)}}\times d}$}
\LeftComment Next, apply self-attention to all sequences together
\State $g=\text{RMSNorm}(f)$\Comment{$\mathbb{R}^{L_x\times d}$}
\State $q^{(i)}_j=\text{RoPE}(\text{Linear}(g^{(i)}_j), j) \hspace{1mm} \forall i,j$\Comment{$\mathbb{R}^d$}
\State $k^{(i)}_j=\text{RoPE}(\text{Linear}(g^{(i)}_j), j) \hspace{1mm} \forall i,j$\Comment{$\mathbb{R}^d$}
\State $v=\text{Linear}(g)$\Comment{$\mathbb{R}^{L_x\times d}$}
\State $g=g+\text{Attention}(q,k,v)$\Comment{$\mathbb{R}^{L_x\times d}$}
\LeftComment Finally, the feedforward layer
\State $g'=\text{RMSNorm}(g)$\Comment{$\mathbb{R}^{L_x\times d}$}
\State $h'=\text{SwiGLU}(g')$\Comment{$\mathbb{R}^{L_x\times \frac{8}{3}d}$}
\State $g'=g + \text{Linear}(h')$\Comment{$\mathbb{R}^{L_x\times d}$}
\State \textbf{return} $g'$
\end{algorithmic}
\end{algorithm}

\begin{algorithm}
\caption{\texttt{encoder} -- encodes a prompt $x$ composed of a sequence-of-sequences}\label{alg:encoder}
\begin{algorithmic}[1]
\Require $\forall i\in\{1..\text{N}\}$
\item[\hspace{1em}\textbullet~$x_\text{inputs}=\{x_\text{seq}, x_\text{plddt}, x_\text{atomb}\}$]
\item[\hspace{1em}\textbullet~pairwise discretized C$\alpha$ distances $D^{(i)}_{m,n}\in\{1..128\}, m\in\{1..L_{x_i}\}, n\in\{1..L_{x_i}\}$]
\State $h_\text{encoder}=\texttt{embed\_inputs}(x_{\text{inputs}})$ \Comment{$\mathbb{R}^{L_x\times d}$}
\For{$l\in{1..n_\text{layers}}$}
\State $h_\text{encoder}=\texttt{encoder\_layer}(h_\text{encoder}, D)$
\EndFor
\State $z_\text{seq} = \text{Linear}(h)$ \Comment{$\mathbb{R}^{L_x\times 28}$}
\State \textbf{return} prompt embedding $h_\text{encoder}$, sequence logits $z_\text{seq}$
\end{algorithmic}
\end{algorithm}

\subsection{Decoder Architecture}

The decoders each decode a single sequence $y$, conditioned on (1) the prompt embedding $h_\text{encoder}$, and (2) an optional query index $q$ indicating which sequence in the prompt to use as the query, if any.

The transformations performed in each decoder are detailed in Algorithm \ref{alg:decoder}; the CLM decoder and MLM decoder only differ in that the former uses a causal mask in the attention operations, and the latter does not. First, the input $y$ is embedded. Next, if a query is present, the embedding of $y$ and the embedding of the query (produced by the encoder) are averaged. Then, $n_\text{layers}$ of decoder layers (Algorithm \ref{alg:decoder_layer}) that are equivariant to the protein order of the prompt are applied. The decoders each have a single output, $z_\text{seq}$, that is used to compute the corresponding loss ($\mathcal{L}_\text{CLM decoder}$ for the CLM decoder and $\mathcal{L}_\text{MLM decoder}$ for the MLM decoder).

\begin{algorithm}
\caption{\texttt{decoder\_layer} (for brevity, this algorithm describes only a single attention head, but can be extended to multiple attention heads in the normal fashion)}\label{alg:decoder_layer}
\begin{algorithmic}[1]
\Require
\item[\hspace{1em}\textbullet~decoder type $T\in\{\text{CLM}, \text{MLM}\}$]
\item[\hspace{1em}\textbullet~encoder prompt embeddings $h^{(i)}_{\text{encoder},j}\in\mathbb{R}^d, i\in\{1..\text{N}\}, j\in\{1..L_{x_i}\}$]
\item[\hspace{1em}\textbullet~decoder sequence embedding $h_{\text{decoder},i}\in\mathbb{R}^d, i\in\{1..L_y\}$]
\item[\hspace{1em}\textbullet~pairwise discretized C$\alpha$ distances $D_{m,n}\in\{1..128\}, m\in\{1..L_{y}\}, n\in\{1..L_{y}\}$]
\item[]
\LeftComment First, apply self-attention with structure-based attention bias
\State $f=\text{RMSNorm}(h_\text{decoder})$ \Comment{$\mathbb{R}^{L_y\times d}$}
\State $q_i=\text{RoPE}(\text{Linear}(f_i), i) \hspace{1mm} \forall i$ \Comment{$\mathbb{R}^d$}
\State $k_i=\text{RoPE}(\text{Linear}(f_i), i) \hspace{1mm} \forall i$ \Comment{$\mathbb{R}^d$}
\State $v=\text{Linear}(f)$\Comment{$\mathbb{R}^{L_y\times d}$}
\LeftComment Compute attention score with structure-based bias
\State $A_{m,n}={q_m}^Tk_n+\texttt{structure\_bias}(D_{m,n})$\item[]\Comment{$\mathbb{R}$}
\If{$T==\text{CLM}$}{
$A=A+\text{CausalMask}(L_y)$\Comment{$\mathbb{R}^{L_y\times L_y}$}
}\EndIf
\State $f=f+\text{softmax}(\frac{A}{\sqrt{d}})v$ \hspace{1mm} \Comment{$\mathbb{R}^{L_y\times d}$}
\LeftComment Next, apply cross-attention to prompt embeddings
\State $g=\text{RMSNorm}(f)$\Comment{$\mathbb{R}^{L_y\times d}$}
\State $q_i=\text{RoPE}(\text{Linear}(g_i), i) \hspace{1mm} \forall i$\Comment{$\mathbb{R}^d$}
\State $k^{(i)}_j=\text{RoPE}(\text{Linear}(h^{(i)}_{\text{encoder},j}), j) \hspace{1mm} \forall i,j$\Comment{$\mathbb{R}^d$}
\State $v=\text{Linear}(h_\text{encoder})$\Comment{$\mathbb{R}^{L_y\times d}$}
\State $g=g+\text{Attention}(q,k,v)$\Comment{$\mathbb{R}^{L_y\times d}$}
\LeftComment Finally, the feedforward layer
\State $g'=\text{RMSNorm}(g)$\Comment{$\mathbb{R}^{L_y\times d}$}
\State $h_\text{decoder}'=\text{SwiGLU}(g')$\Comment{$\mathbb{R}^{L_y\times \frac{8}{3}d}$}
\State $g'=g + \text{Linear}(h_\text{decoder}')$\Comment{$\mathbb{R}^{L_y\times d}$}
\State \textbf{return} $g'$
\end{algorithmic}
\end{algorithm}

\begin{algorithm}
\caption{\texttt{decoder} -- decodes a single sequence $y$ conditioned on prompt embeddings $h_\text{encoder}$}\label{alg:decoder}
\begin{algorithmic}[1]
\Require
\item[\hspace{1em}\textbullet~decoder type $T\in\{\text{CLM}, \text{MLM}\}$]
\item[\hspace{1em}\textbullet~optional query index $q\in\{0...\text{N}\}$]
\item[\hspace{1em}\textbullet~encoder prompt embeddings $h^{(i)}_{\text{encoder},j}\in\mathbb{R}^d, i\in\{1..\text{N}\}, j\in\{1..L_{x_i}\}$]
\item[\hspace{1em}\textbullet~decoder inputs $y_\text{inputs}=\{y_\text{seq}, y_\text{plddt}, y_\text{atomb}\}$]
\item[\hspace{1em}\textbullet~pairwise discretized C$\alpha$ distances $D_{m,n}\in\{1..128\}, m\in\{1..L_{y}\}, n\in\{1..L_{y}\}$]
\item[]
\State $h_\text{decoder}=\texttt{embed\_inputs}(y_{\text{inputs}})$ \Comment{$\mathbb{R}^{L_y\times d}$}
\LeftComment Embed the query if there is one
\If{$q \neq 0$}
\State $h_{\text{decoder},i}=\frac{1}{2}(h_{\text{decoder},i}+h^{(q)}_{\text{encoder},i})\forall i$\Comment{$\mathbb{R}^d$}
\EndIf
\For{$l\in{1..n_\text{layers}}$}
\State $h_\text{decoder}=\texttt{decoder\_layer}(T, h_\text{encoder}, h_\text{decoder}, D)$\Comment{$\mathbb{R}^{L_y\times d}$}
\EndFor
\State $z_\text{seq} = \text{Linear}(h_\text{decoder})$ \Comment{$\mathbb{R}^{L_y\times 28}$}
\State \textbf{return} sequence logits $z_\text{seq}$
\end{algorithmic}
\end{algorithm}

\clearpage
\subsection{Miscellaneous}\label{appendix:mic_architecture}

This section details additional aspects of PoET-2's architecture that are related to capabilities that are \textit{not} utilized in this paper's experiments.

\subsubsection{3Di Token Prediction}

In addition to predicting each protein's amino acid sequence, PoET-2 is also trained to predict each protein's 3Di structure token sequence \cite{foldseek} using the cross entropy loss. The 3Di structure tokens are only predicted when the predicted pLDDT of the residue is at least $70$. The amino acid and 3Di structure token losses have equal weight i.e. the total loss is simply the sum of the two losses.

\subsubsection{Conditioning on target homology}

PoET-2 is also trained to generate sequences that must be within a specified sequence identity range of a protein in the prompt, referred to as the query protein. The query can be any protein in the prompt whose sequence is completely known (i.e. contains no unknown or masked amino acids). This generation mode is called "target homology". When using this generation mode, the structure of the generated protein does not have to contain any structural elements specified in the query protein.

The target homology generation mode is implemented with two modifications to the architecture discussed so far:

\begin{enumerate}
    \item The input embedding is augmented with another feature, $x_\text{seqid}$. This feature represents the sequence identity range as two values $\in[0,1]$ indicating the range's lower and upper bounds, and is repeated across all residues. Thus, it has shape $L\times 2$. If the target homology generation mode is inactive (e.g. as in the encoder) or not being used in the decoder, these two sequence identity values are both set to 0. To prepare $x_\text{seqid}$ for input embedding, it is processed similarly to other features like $x_\text{plddt}$ and $x_\text{atomb}$: any values at positions intended to be masked are set to zero, and then a corresponding binary mask (1 for observed/valid, 0 for masked/invalid) is concatenated as a third channel to these two sequence identity values. This augmented 3-channel tensor for $x_\text{seqid}$ is then linearly projected to the model's hidden dimension and subsequently summed with the embeddings of other input features, as detailed in Algorithm \ref{alg:embed_inputs}.
    \item Since the generated sequence does not necessarily have the same length as the query, we cannot simply combine the embeddings of the generated sequence and the embeddings of the query by summing the embeddings of all corresponding residues as in Line 3 of Algorithm \ref{alg:decoder}, since the correspondence is unknown. Instead, we sum the embedding of each residue of the generated sequence with the embedding of the first residue of the query sequence. That is, when the target homology generation mode is used, we replace Line 3 of Algorithm \ref{alg:decoder} with $h_{\text{decoder},i}=\frac{1}{2}(h_{\text{decoder},i}+h^{(q)}_{\text{encoder},1}) \forall i$.
\end{enumerate}

\subsubsection{Decoding queries of unknown length}

Lastly, PoET-2 is trained to decode query sequences with contiguous segments of unknown length. These segments are represented with the gap token (\texttt{-}) mentioned in \S\ref{model_inputs}; the gap token indicates zero or more residues with unknown identity. For example, the query sequence $\texttt{\$MK-IP*}$ indicates that PoET-2 must generate a sequence that starts with the two amino acids M and K, then has 0 of more amino acids, and then ends with the amino acids I and P.

In order to decode such sequences, PoET-2 uses a special decoding scheme called the "insertion decoding scheme". The purpose of this decoding scheme is to align the residues of the query sequence and the generated sequence so that the embeddings of their corresponding residues can be summed in Line 3 of Algorithm \ref{alg:decoder}.

In the standard decoding scheme, the alignment between the query sequence and the sequence being generated by the model may be ambiguous when gap tokens are present in the query. For example, suppose that the query sequence is \texttt{\$-A}. In a normal decoding scheme, if the model predicts that the token following the start token is the token A, it is ambiguous if that token should be aligned with the gap token to indicate an insertion, or aligned with the token A to indicate that there are no insertions. In order to address this ambiguity, we train the decoders to instead output the gap token when generating a token that is unmasked in the query sequence. Continuing the example, if the model wants to generate the token A as part of an insertion aligned with the gap token in the query sequence, then the model should simply output the token A as usual. However, if the model wants to generate the token A and have it be aligned with the token A at the third position in the query sequence, then the model should output the gap token. In this case, the gap token in the query sequence represents no insertions.

\begin{figure}[htbp]
  \centering
  \includegraphics[width=\linewidth]{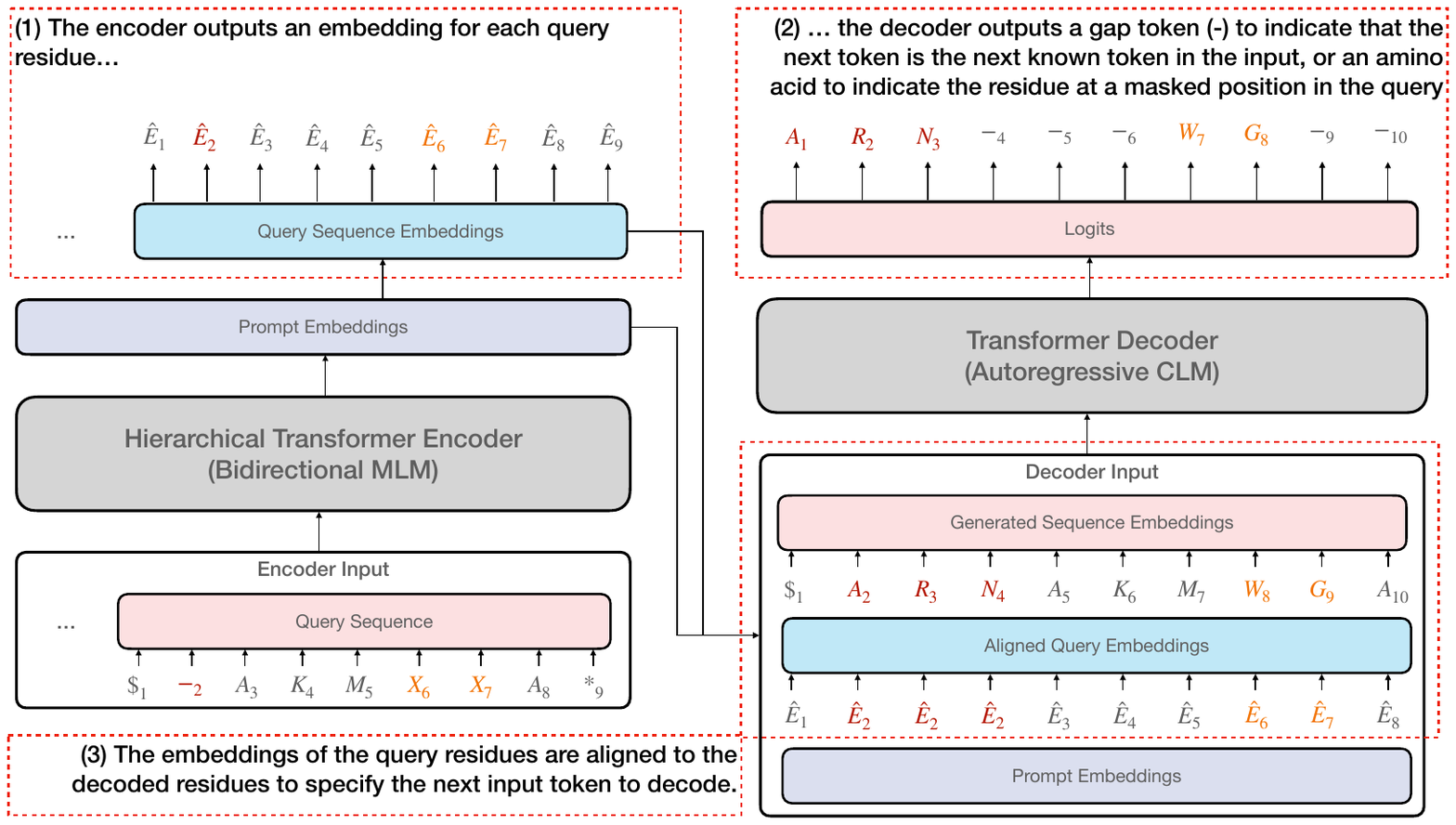}
  \caption{Visualization of insertion decoding scheme.}
  \label{fig:insertion_decoding_scheme}
\end{figure}

The insertion decoding scheme is visualized in Figure \ref{fig:insertion_decoding_scheme}. Using the alignment provided by the insertion decoding scheme, Line 3 of Algorithm \ref{alg:decoder} can then be modified to be $h_{\text{decoder},i}=\frac{1}{2}(h_{\text{decoder},i} + h^{(q)}_{\text{encoder},\text{alignment}_i}) \forall i$, where $\text{alignment}_i$ indicates the index of the residue of the query that the $i$th residue of the generated sequence is aligned to.

Although the insertion decoding scheme is primarily required for the CLM decoder, we apply the insertion decoding scheme to the MLM decoder as well by adjusting the outputs tokens in a similar manner i.e. since the MLM decoder is trained to unmask the token at the current position (as opposed to next token prediction for the CLM decoder), if the token at the current position is unmasked, the MLM decoder is trained to output the gap token.

\section{PoET-2 Training Details}\label{appendix:poet_2_training_details}

\subsection{Training data}

\paragraph{Sequence data} PoET-2 is trained on sets of homologous sequences. Sets of homologous sequences are found and sampled using the same procedure used by PoET-1 \cite{poet1}. Summarizing briefly, the sets of homologous sequences are found by using Diamond \cite{diamond} to search UniRef50 in an all against all search using the following command:

\texttt{diamond blastp -q uniref50.fasta -d diamond/uniref50 -f 6 --header -k 200000 --max-hsps 1 -e 0.001 -p 96 -o output.tab}

Sets of homologous sequences are sampled with weight proportional to the inverse of the size of the set.

The main differences between the training data for PoET-2 and PoET-1 are as follows:

\begin{itemize}
    \item UniRef Version 2304 is used instead of UniRef Version 2103.
    \item All sets of homologous sequences are used, rather than only sets with at least 10 members.
\end{itemize}

As a result of these differences, PoET-2 is trained on 62 million sets, as opposed to 29 million sets for PoET-1.

\paragraph{Structure data} PoET-2 is trained only on predicted structures from AFDB \cite{afdb,afdb-2}, and no experimentally solved structures. Sequences in the training data are associated with structures in AFDB using the UniRef100 sequence identifier. When the structure of a protein is used as input to PoET-2, if there is a conflict between the sequence in UniRef and AFDB (e.g. due to changes in UniRef), the sequence in AFDB is used. On the order of approximately half of sequences in the training data can be associated with a predicted structure using this methodology.

\subsection{Noise schedule}

For sequence inputs,

\begin{itemize}
    \item Context sequence tokens are masked with a random masking rate chosen uniformly from 0\%-30\%.
    \item Query sequence tokens are randomly masked with a random masking rate chosen uniformly from 0\%-100\%.
    \item Decoder sequence tokens are masked with a random masking rate chosen uniformly from 0\%-30\%.
\end{itemize}

For structure inputs, the pLDDT and atomic backbone coordinates of the N, C$\alpha$, and C atoms are masked with a random masking rate chosen uniformly from 0\%-100\%.

For both sequence and structure inputs, with probability 

\begin{itemize}
    \item 50\%, masking is performed randomly per residue.
    \item 25\%, random contiguous spans of length L are masked, where L is drawn from the distribution \texttt{Poisson}(3) + 1.
    \item 25\%, N random contiguous spans are masked, where N is drawn randomly from \texttt{Poisson}(2.5) half the time and from \texttt{Poisson}(13) the other half of the time.
\end{itemize}

\subsection{Optimizer and learning rate schedule}

PoET-2 is trained with the same optimizer and learning rate schedule as PoET-1 \cite{poet1}. Namely, the optimizer is Adafactor \cite{adafactor}, and the learning rate schedule consists of a linear warmup over the first $4000$ steps to a peak learning rate of $1e-2$, and then a square root decay over the remaining training steps.

\subsection{Compute requirements}

PoET-2 is trained for 3 million steps on 8 x A100 GPUs with 40GB VRAM each. A batch size of 45056 tokens is used per GPU with gradient accumulation over two steps, for an effective batch size of 90112 tokens per GPU. The total training time on this hardware is approximately 2.5 months.

\clearpage
\section{Zero-shot variant effect prediction}\label{appendix:zero_shot}

\subsection{Prompt engineering}\label{appendix:prompt_engineering_zero_shot}

Recall that the prompt consists of two optional components, a context containing proteins from the protein family of interest, and a query containing explicit sequence and/or structure constraints. Our goal in prompt engineering is to design prompts that enable us to accurately predict the properties of variants of a WT sequence.

To determine the set of proteins to use in the context, we use the same method as PoET-1 \cite{poet1}. We first identify proteins in the protein family by searching for sequence homologs of WT in UniRef100 \cite{uniprot} using the ColabFold MSA protocol \cite{colabfold}. We then select the proteins to include in the context of a prompt by sampling a representative subset of the sequence homologs using the method from Hopf et. al \cite{potts}.

Building off of this approach, we employ two prompt engineering strategies to improve predictions:

\begin{itemize}
    \item Following the approach of PoET-1 \cite{poet1}, we ensemble over different prompts, where the context of each prompt contains a different subsample of sequence homologs. The ensemble prediction is simply the average of the individual predictions. Furthermore, for each context, we use different values for the context length (i.e. number of tokens or amino acids in the context) and maximum similarity of a sequence in the context to the WT sequence. The exact values of these parameters used in the ensemble for DMS and clinical datasets are specified in the sections below.
    \item Utilizing PoET-2's multimodal capabilities, we explore two methods for incorporating structure in the prompt. The first method incorporates structure in the context by associating sequences in the context with their predicted structure in AFDB \cite{afdb, afdb-2}, if the sequence exists in AFDB. The second method incorporates structure by adding a query to the prompt that contains the structure (but not the sequence) of WT. The use of this "inverse-folding" query instructs PoET-2 to score the likelihood that a variant sequence will fold into the same structure as WT. Although it is not necessarily desirable for a variant to adopt the same structure as WT, the inverse-folding likelihood has been shown to be predictive of protein fitness, particularly for stability related properties \cite{proteingym}.
    
    Not all methods of incorporating structure in the prompt are always helpful; the best method for doing so on the ProteinGym DMS and clinical substitutions benchmarks are ablated and identified in the following sections. Also, note that the query-based approach is not used for indel variants because indel variants have different lengths from WT and thus cannot adopt the same tertiary structure as WT.
\end{itemize}

\subsubsection{Deep mutational scanning datasets}

\paragraph{Ensembling over context length and maximum similarity} Following PoET-1 \cite{poet1}, we ensemble over all combinations of values for context length $\in\{6144, 12288, 24576\}$ and maximum similarity $\in\{1.0, 0.95, 0.90, 0.70, 0.50\}$, resulting in 15 combinations in total.

\paragraph{Incorporating structure in the prompt}

Table \ref{tab:unsupervised_dms_prompt_engineering} shows the performance of various strategies for incorporating or not incorporating structure in the prompt, and the performance of ensembling different strategies. First, we analyze the effect of different strategies on the substitutions benchmark. When not incorporating structure at all (Strategy A), thus using the same prompting strategy as PoET-1, PoET-2 performs marginally better than PoET-1 ($\Delta\rho=0.005$; PoET-1 reported in Table \ref{tab:dms_unsupervised}). Both including the structure in the context (Strategy B), and in the query (Strategy C) improves performance, with the latter strategy offering a larger improvement. Interestingly, combining the two approaches (Strategy D) performs only about the same or slightly worse than Strategy C ($\Delta\rho=-0.002$).

Strategy I, which ensembles all of the above strategies (A-D), improves performance further by $\Delta\rho=0.009$ vs Strategy C, the best individual strategy. However, we find that Strategy H, which only ensembles Strategies B and D and excludes the strategies that include only sequence in the context, performs similarly to Strategy I, with negligible performance loss. Therefore, we recommend the use of Strategy H, and use this strategy in performance comparisons with other models.

For indel variants, we observe a small improvement in performance by incorporating structure in the context (Strategy B) versus not (Strategy A). There is little or no benefit to ensembling these two strategies (Strategy E). Since a query cannot be used, the other strategies are not applicable. Therefore, we employ Strategy B when comparing PoET-2 to other models.

\subsubsection{Clinical datasets}

\paragraph{Ensembling over context length and maximum similarity} Following PoET-1 \cite{poet1_proteingym_pr}, we use a context length of $49152$ and ensemble over different values for maximum similarity $\in\{1.0, 0.95, 0.90, 0.70, 0.50\}$, resulting in 5 combinations in total. Note that the ensembling parameters for clinical datasets is less well studied than for DMS datasets, and there are likely better parameters for ensembling.

\paragraph{Incorporating structure in the prompt}

\begin{table}[htbp]
\centering
\caption{Performance (AUROC) of different strategies for including structure in the prompt on the zero-shot clinical benchmarks.}
\label{tab:unsupervised_clinical_prompt_engineering}
\begin{tabular}{c|cccc}
\toprule
\multirow{2}{*}{\textbf{Strategy}} & \multicolumn{2}{c}{\textbf{Prompt}} & \multirow{2}{*}{\textbf{Substitutions}} & \multirow{2}{*}{\textbf{Indels}} \\
\cmidrule{2-3}
& \textbf{Context Modalities} & \textbf{Query} & \\
\midrule
A & Sequence & None & 0.92544 & 0.94826 \\
B & Sequence and Structure & None & \textbf{0.92624} & \textbf{0.95278} \\
C & Sequence & Structure of WT & 0.91505 & N/A \\
D & Sequence and Structure & Structure of WT & 0.91292 & N/A \\
\midrule
E & \multicolumn{2}{c}{\makecell[c]{Ensemble of A and B \\ (Different contexts, No query)}} & \textbf{0.92789} & \textbf{0.95179} \\
\addlinespace[1.5pt]
\cdashline{1-5}
\addlinespace[1.5pt]
F & \multicolumn{2}{c}{\makecell[c]{Ensemble of C and D \\ (Different contexts, With query)}} & 0.91599 & N/A \\
\addlinespace[1.5pt]
\cdashline{1-5}
\addlinespace[1.5pt]
G & \multicolumn{2}{c}{\makecell[c]{Ensemble of A and C \\ (Sequence only context, Different queries)}} & 0.92481 & N/A \\
\addlinespace[1.5pt]
\cdashline{1-5}
\addlinespace[1.5pt]
H & \multicolumn{2}{c}{\makecell[c]{Ensemble of B and D \\ (Sequence and structure context, Different queries)}} & 0.92481 & N/A \\
\addlinespace[1.5pt]
\cdashline{1-5}
\addlinespace[1.5pt]
I & \multicolumn{2}{c}{Ensemble of A, B, C, and D} & 0.92572 & N/A \\
\bottomrule
\end{tabular}%
\end{table}

Table \ref{tab:unsupervised_clinical_prompt_engineering} shows the performance of various strategies for incorporating or not incorporating structure in the prompt, and the performance of ensembling different strategies. First, we analyze the effect of different strategies on the substitutions benchmark. Incorporating structure in the context (Strategy B) offers a very minor and not statistically significant improvement versus not incorporating structure in the context (Strategy A). Incorporating structure via the query (Strategy C), however, has a negative effect. Therefore, we do not consider strategies that use a query further.

Ensembling Strategies A and B (Strategy E) has a small positive effect over not ensembling, although due to the small effect size, it is unclear if the effect is simply due to ensembling more prompts, or due to ensembling different prompt strategies. Nevertheless, since Strategy E performs best, we use it in comparisons with other models.

For indel variants, we observe similar trends among applicable strategies (those not using a query), with some improvement observed for incorporating structure in the context, but variations in performance being fairly minor. Although Strategy E slightly underperforms Strategy B by a non-statistically significant amount, we employ Strategy E in comparisons with other models for consistency with the strategy used for the substitutions benchmark.

\subsection{Length adjusted log likelihood (ratio)}\label{appendix:length_adjusted_llr}

Sequence likelihoods from autoregressive models trained with next-token prediction losses and teacher forcing can exhibit miscalibrated stop token probabilities that bias them towards shorter sequences. This likely arises because the loss function only operates at the token level -- it does not strongly penalize an early stop token as long as the early stop token is predicted to be relatively unlikely compared to other tokens.

\begin{figure}[htbp]
  \centering
  \includegraphics[width=0.7\linewidth]{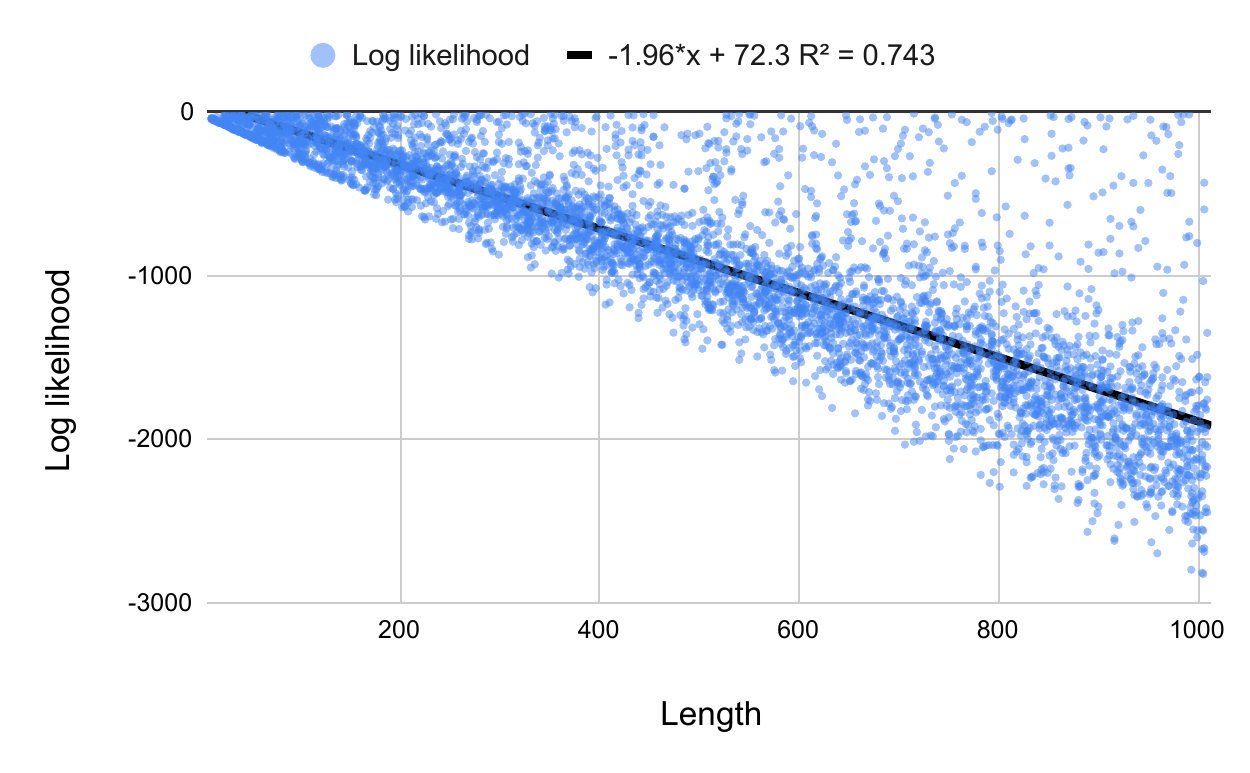}
  \caption{Plot of log likelihood vs length for random UniRef50 protein families.}
  \label{fig:log_likelihood_vs_length}
\end{figure}

This bias towards shorter sequences can be problematic when scoring indel variants with likelihoods, as indel variants can differ in length from WT and each other. To compensate for this, when scoring indel variants, we apply an adjustment to the log likelihood that favors longer sequences over shorter sequences. We find that on a sample of random protein families from UniRef50, the log likelihood decreases roughly linearly with sequence length, with a slope of $-1.96$ (Figure \ref{fig:log_likelihood_vs_length}).

Therefore, we compute the length adjusted log likelihood as follows:

\begin{equation}
    \text{adjusted log likelihood} = \text{log likelihood} + \alpha \times \text{sequence length}
\end{equation}

where $\alpha=1.96$ is the length adjustment factor. To compute the adjusted log likelihood \textit{ratio} for scoring variants, we simply use the adjusted log likelihood instead of the regular log likelihood.

\begin{figure}[htbp]
  \centering
  \includegraphics[width=0.5\linewidth]{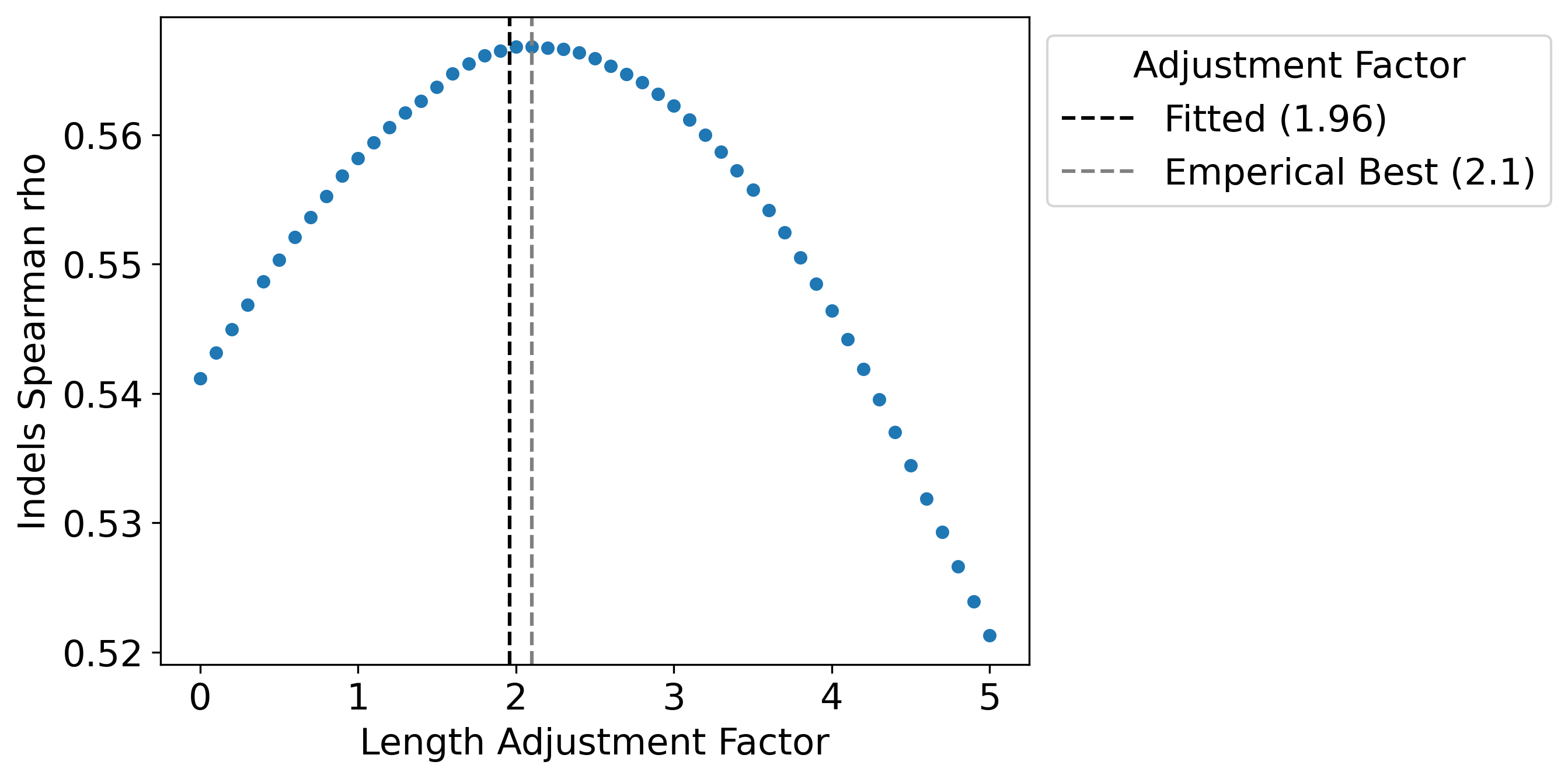}
  \caption{Plot of DMS indels performance ($\rho$) vs length adjustment factor.}
  \label{fig:dms_indels_vs_adjustment_factor}
\end{figure}

On the DMS indels benchmark, we find that using the length adjusted likelihood ratio with $\alpha=1.96$ improves performance (Figure \ref{fig:dms_indels_vs_adjustment_factor}). The length adjustment factor $\alpha=1.96$ is near optimal, with the empirical best adjustment factor being $2.1$. 

\begin{figure}[htbp]
  \centering
  \begin{subfigure}{0.48\textwidth}
    \centering
    \includegraphics[width=\linewidth]{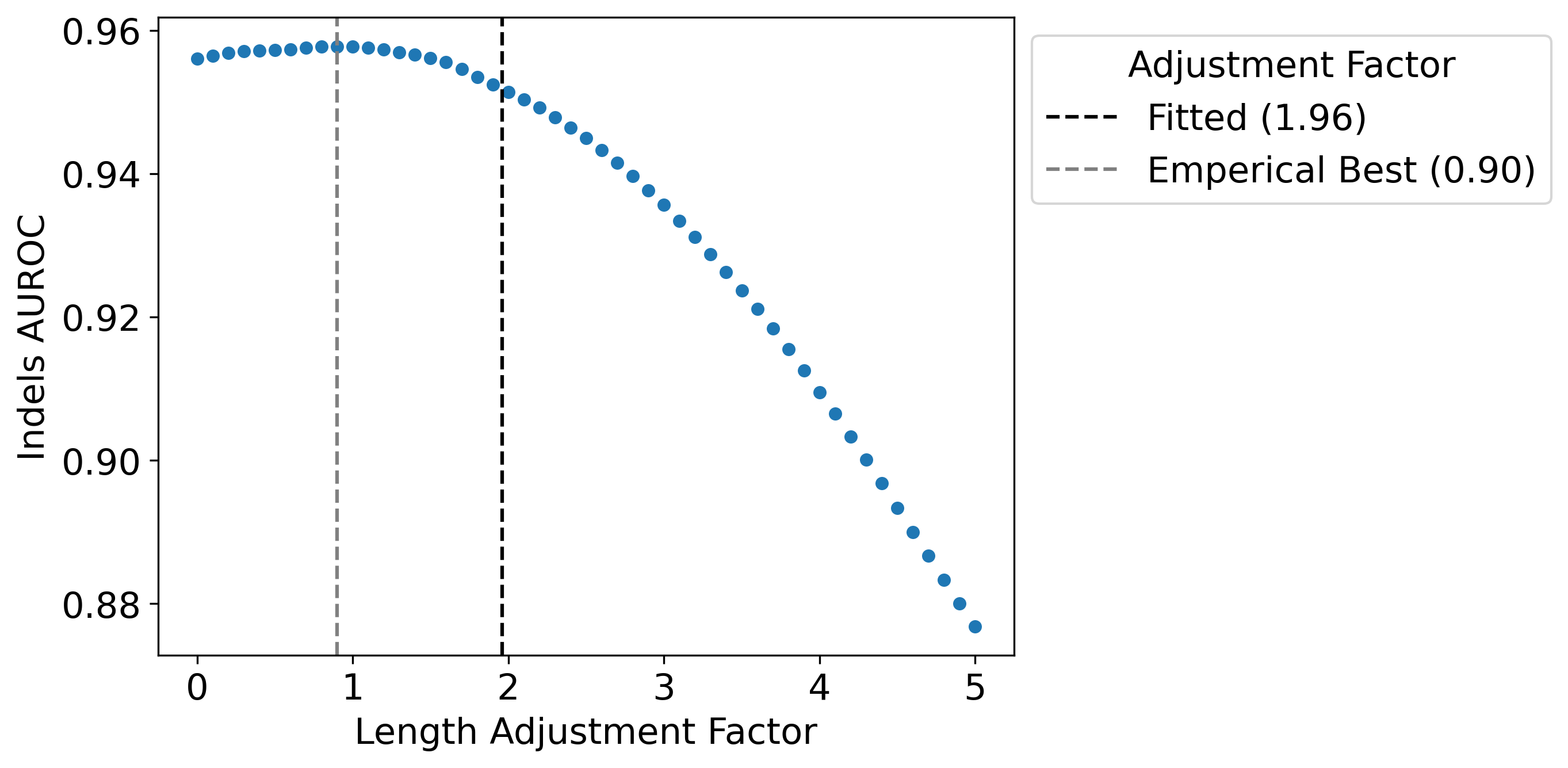}
  \end{subfigure}
  \hfill
  \begin{subfigure}{0.48\textwidth}
    \centering
    \includegraphics[width=\linewidth]{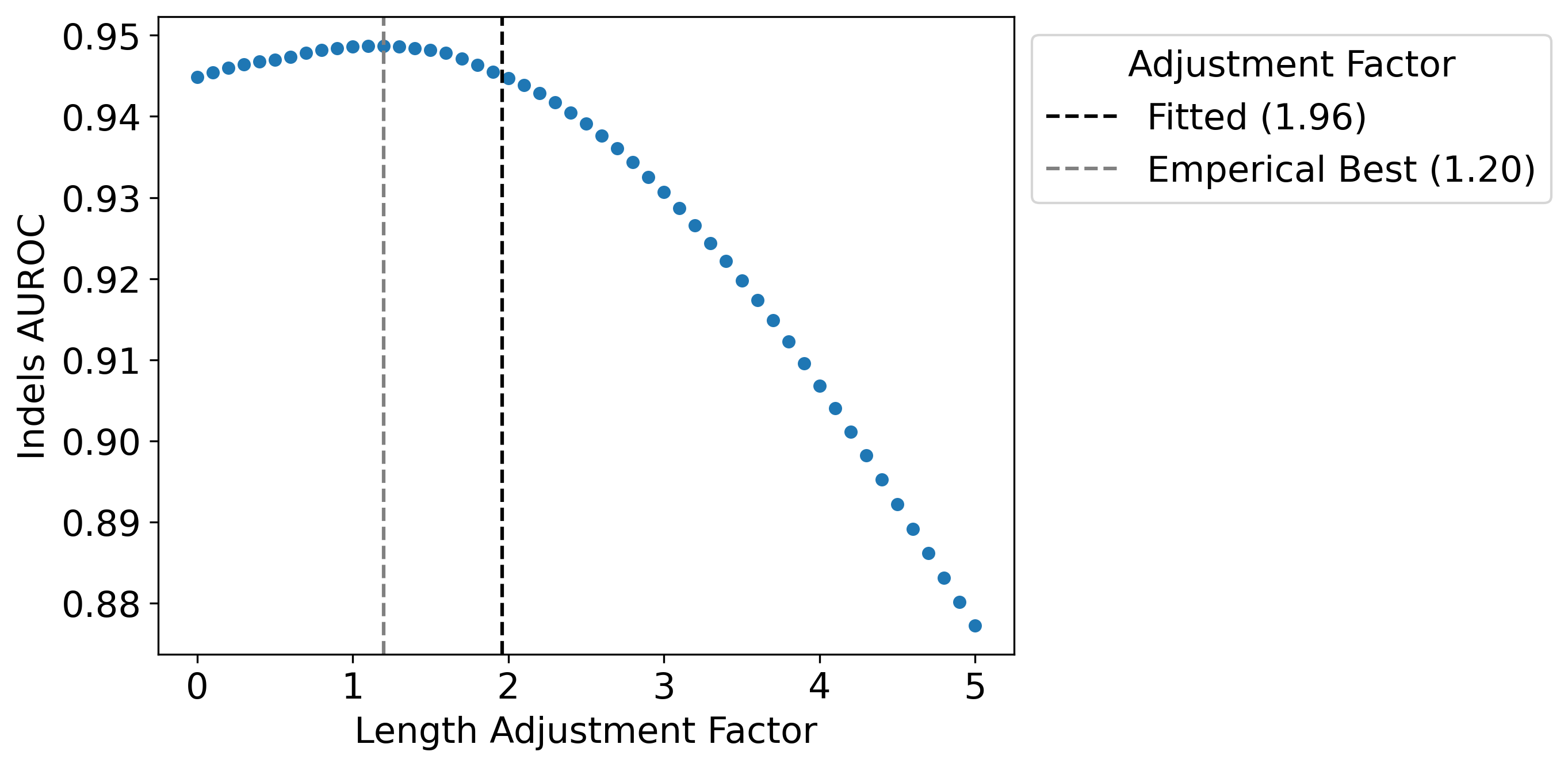}
  \end{subfigure}
  \caption{Plot of clinical indels performance (AUROC) versus length adjustment factor; results for ProteinGymV1 (left) and ProteinGymV1.3 (right).}
  \label{fig:clinical_indels_vs_adjustment_factor}
\end{figure}

On the other hand, on the clinical indels benchmark, we find that using the length adjusted likelihood ratio with $\alpha=1.96$ slightly harms performance (Figure \ref{fig:clinical_indels_vs_adjustment_factor}). However, the optimal adjustment factor is non-zero (between $0.90$ and $1.20$ depending on the benchmark version), indicating that some length adjustment is generally ideal for zero-shot fitness prediction, regardless of the specific task. Given that the relation between log likelihood and length is not completely linear, and that the length only explains \textasciitilde 75\% of the variance in the log likelihood (Figure \ref{fig:log_likelihood_vs_length}), there may be better ways to adjust the log likelihood using factors other than the length e.g. factors that may be dependent on the specific protein family of interest. We leave exploration of this to future work.

As our experiments show that adjusting log likelihoods for length is generally useful, even if $\alpha=1.96$ is not necessarily optimal, we always use the length adjusted log likelihood when comparing the performance of PoET-2 to other models.

\subsection{Model ensembles}\label{appendix:zero_shot_ensembles}

We compute the zero-shot score for the ensemble model combining PoET-2 and VenusREM by computing a weighted average of the score from PoET-2 and the score from VenusREM:

\begin{equation}
    \text{Ensemble Score} = w\times(\text{PoET-2 score}) + (1 - w)\times(\text{VenusREM score})
\end{equation}

where $w\in[0,1]$. To select the weight $w$, we evaluate the performance of $21$ values of $w$ regularly spaced in the interval $[0,1]$ (inclusive; increments of $0.05$) on ProteinNPT's \cite{proteinnpt} validation of set $8$ datasets:

\texttt{\detokenize{BLAT_ECOLX_Jacquier_2013, CALM1_HUMAN_Weile_2017, DYR_ECOLI_Thompson_2019, DLG4_RAT_McLaughlin_2012, P53_HUMAN_Giacomelli_2018_WT_Nutlin, REV_HV1H2_Fernandes_2016, RL40A_YEAST_Roscoe_2013, TAT_HV1BR_Fernandes_2016}}

On this validation set, we find that the optimal value of $w$ is simply $0.5$, corresponding to a simple average.

\subsection{Detailed results}\label{appendix:zero_shot_detailed_results}

The following tables detail results on the performance and standard error of models on the DMS zero-shot substitutions and indels benchmarks, broken by different metrics, and assay and protein subgroups.

\begin{itemize}
    \item \textbf{Table \ref{tab:dms_subs_all_vs_poet_2}}: Overall performance (Spearman, AUC, MCC, NDCG, Recall) on substitutions benchmark, with standard error of difference to PoET-2.
    \item \textbf{Table \ref{tab:dms_subs_all_vs_poet_2_venus_rem}}: Overall performance (Spearman, AUC, MCC, NDCG, Recall) on substitutions benchmark, with standard error of difference to PoET-2 + VenusREM.
    \item \textbf{Table \ref{tab:dms_indels_all_vs_poet_2}}: Overall performance (Spearman, AUC, MCC, NDCG, Recall) on indels benchmark, with standard error of difference to PoET-2.
    \item \textbf{Table \ref{tab:dms_subs_assay_type}}: Performance (Spearman) on substitutions benchmark broken down by assay type, with standard error of difference to PoET-2.
    \item \textbf{Table \ref{tab:dms_subs_msa_depth}}: Performance (Spearman) on substitutions benchmark broken down by MSA depth, with standard error of difference to PoET-2.
    \item \textbf{Table \ref{tab:dms_subs_taxon}}: Performance (Spearman) on substitutions benchmark broken down by taxonomy, with standard error of difference to PoET-2.
    \item \textbf{Table \ref{tab:dms_subs_mutation_depth}}: Performance (Spearman) on substitutions benchmark broken down by mutation depth, with standard error of difference to PoET-2.
\end{itemize}

\begin{table}[htbp]
\caption{Performance on zero-shot DMS substitutions benchmark. Standard error of difference to PoET-2 in parentheses.}
\label{tab:dms_subs_all_vs_poet_2}
\centering
\resizebox{\textwidth}{!}{
\begin{tabular}{lccccc}
\toprule
\multirow{2}{*}{\textbf{Model}} & \multicolumn{5}{c}{\textbf{Metric}} \\
\cmidrule(lr){2-6}
& \textbf{Spearman} & \textbf{AUC} & \textbf{MCC} & \textbf{NDCG} & \textbf{Recall} \\
\midrule
ESM-2 & 0.415 (0.013) & 0.729 (0.007) & 0.328 (0.010) & 0.742 (0.007) & 0.217 (0.007) \\
ESM C & 0.407 (0.014) & 0.727 (0.007) & 0.323 (0.011) & 0.742 (0.007) & 0.213 (0.006) \\
ProGen2 M & 0.379 (0.009) & 0.711 (0.005) & 0.299 (0.007) & 0.747 (0.006) & 0.203 (0.006) \\
ProGen2 XL & 0.390 (0.009) & 0.717 (0.005) & 0.306 (0.008) & 0.764 (0.004) & 0.198 (0.005) \\
\midrule
SaProt & 0.457 (0.007) & 0.751 (0.004) & 0.358 (0.007) & 0.764 (0.005) & 0.232 (0.006) \\
ESM-3 Open & 0.467 (0.005) & 0.756 (0.003) & 0.368 (0.005) & 0.773 (0.004) & 0.241 (0.006) \\
ProSST & 0.508 (0.008) & 0.777 (0.004) & 0.399 (0.007) & 0.752 (0.007) & 0.235 (0.009) \\
\midrule
MSA Transformer & 0.431 (0.009) & 0.736 (0.005) & 0.338 (0.007) & 0.774 (0.004) & 0.225 (0.005) \\
TranceptEVE L & 0.456 (0.006) & 0.751 (0.003) & 0.356 (0.005) & 0.783 (0.003) & 0.231 (0.005) \\
GEMME & 0.455 (0.009) & 0.749 (0.005) & 0.352 (0.007) & 0.773 (0.003) & 0.212 (0.004) \\
PoET-1 & 0.470 (0.004) & 0.759 (0.002) & 0.367 (0.004) & 0.779 (0.002) & 0.226 (0.003) \\
\midrule
S3F-MSA & 0.496 (0.006) & 0.771 (0.003) & 0.388 (0.005) & 0.788 (0.002) & 0.244 (0.004) \\
VenusREM & 0.519 (0.007) & 0.783 (0.003) & 0.405 (0.006) & 0.766 (0.006) & 0.243 (0.009) \\
PoET-2 & 0.500 (0.000) & 0.773 (0.000) & 0.391 (0.000) & 0.786 (0.000) & 0.238 (0.000) \\
\cdashline{1-6}
PoET-2 + VenusREM & \textbf{0.543 (0.005)} & \textbf{0.796 (0.002)} & \textbf{0.423 (0.004)} & \textbf{0.791 (0.004)} & \textbf{0.254 (0.006)} \\
\bottomrule
\end{tabular}
}
\end{table}

\begin{table}[htbp]
\caption{Performance on zero-shot DMS substitutions benchmark. Standard error of difference to PoET-2 + VenusREM in parentheses.}
\label{tab:dms_subs_all_vs_poet_2_venus_rem}
\centering
\resizebox{\textwidth}{!}{
\begin{tabular}{lccccc}
\toprule
\multirow{2}{*}{\textbf{Model}} & \multicolumn{5}{c}{\textbf{Metric}} \\
\cmidrule(lr){2-6}
& \textbf{Spearman} & \textbf{AUC} & \textbf{MCC} & \textbf{NDCG} & \textbf{Recall} \\
\midrule
ESM-2 & 0.415 (0.014) & 0.729 (0.007) & 0.328 (0.011) & 0.742 (0.007) & 0.217 (0.007) \\
ESM C & 0.407 (0.014) & 0.727 (0.008) & 0.323 (0.012) & 0.742 (0.007) & 0.213 (0.007) \\
ProGen2 M & 0.379 (0.010) & 0.711 (0.005) & 0.299 (0.008) & 0.747 (0.007) & 0.203 (0.008) \\
ProGen2 XL & 0.390 (0.010) & 0.717 (0.005) & 0.306 (0.009) & 0.764 (0.006) & 0.198 (0.007) \\
\midrule
SaProt & 0.457 (0.008) & 0.751 (0.005) & 0.358 (0.007) & 0.764 (0.005) & 0.232 (0.006) \\
ESM-3 Open & 0.467 (0.007) & 0.756 (0.004) & 0.368 (0.006) & 0.773 (0.004) & 0.241 (0.006) \\
ProSST & 0.508 (0.005) & 0.777 (0.003) & 0.399 (0.004) & 0.752 (0.006) & 0.235 (0.006) \\
\midrule
MSA Transformer & 0.431 (0.010) & 0.736 (0.005) & 0.338 (0.008) & 0.774 (0.006) & 0.225 (0.007) \\
TranceptEVE L & 0.456 (0.007) & 0.751 (0.004) & 0.356 (0.006) & 0.783 (0.004) & 0.231 (0.006) \\
GEMME & 0.455 (0.009) & 0.749 (0.005) & 0.352 (0.010) & 0.773 (0.005) & 0.212 (0.007) \\
PoET-1 & 0.470 (0.006) & 0.759 (0.003) & 0.367 (0.006) & 0.779 (0.004) & 0.226 (0.005) \\
\midrule
S3F-MSA & 0.496 (0.006) & 0.771 (0.003) & 0.388 (0.007) & 0.788 (0.003) & 0.244 (0.006) \\
VenusREM & 0.519 (0.003) & 0.783 (0.002) & 0.405 (0.003) & 0.766 (0.004) & 0.243 (0.004) \\
PoET-2 & 0.500 (0.004) & 0.773 (0.002) & 0.391 (0.004) & 0.786 (0.004) & 0.238 (0.006) \\
\cdashline{1-6}
PoET-2 + VenusREM & \textbf{0.543 (0.000)} & \textbf{0.796 (0.000)} & \textbf{0.423 (0.000)} & \textbf{0.791 (0.000)} & \textbf{0.254 (0.000)} \\
\bottomrule
\end{tabular}
}
\end{table}

\begin{table}[htbp]
\caption{Performance on zero-shot DMS indels benchmark. Standard error of difference to PoET-2 in parentheses.}
\label{tab:dms_indels_all_vs_poet_2}
\centering
\resizebox{\textwidth}{!}{
\begin{tabular}{lccccc}
\toprule
\multirow{2}{*}{\textbf{Model}} & \multicolumn{5}{c}{\textbf{Metric}} \\
\cmidrule(lr){2-6}
& \textbf{Spearman} & \textbf{AUC} & \textbf{MCC} & \textbf{NDCG} & \textbf{Recall} \\
\midrule
ProGen2 M & 0.463 (0.037) & 0.770 (0.021) & 0.370 (0.031) & 0.757 (0.018) & 0.305 (0.017) \\
ProGen2 XL & 0.427 (0.022) & 0.747 (0.010) & 0.323 (0.019) & 0.749 (0.015) & 0.297 (0.012) \\
\midrule
TranceptEVE L & 0.410 (0.020) & 0.749 (0.011) & 0.348 (0.020) & 0.725 (0.013) & 0.258 (0.014) \\
PoET-1 & 0.515 (0.006) & 0.803 (0.005) & 0.434 (0.011) & 0.763 (0.006) & 0.310 (0.010) \\
\midrule
PoET-2 & \textbf{0.567 (0.000)} & \textbf{0.831 (0.000)} & \textbf{0.478 (0.000)} & \textbf{0.795 (0.000)} & \textbf{0.340 (0.000)} \\
\bottomrule
\end{tabular}
}
\end{table}

\begin{table}[htbp]
\caption{Performance (Spearman $\rho$) on zero-shot DMS substitutions benchmark broken down by assay type. Standard error of difference to PoET-2 in parentheses.}
\label{tab:dms_subs_assay_type}
\centering
\resizebox{\textwidth}{!}{
\begin{tabular}{lccccc}
\toprule
\multirow{2}{*}{\textbf{Model}} & \multicolumn{5}{c}{\textbf{Substitutions By Assay Type}} \\
\cmidrule(lr){2-6}
& \textbf{Activity} & \textbf{Binding} & \textbf{Expression} & \makecell[c]{\textbf{Organismal}\\\textbf{Fitness}} & \textbf{Stability} \\
\midrule
ESM-2 & 0.429 (0.028) & 0.336 (0.043) & 0.417 (0.030) & 0.368 (0.024) & 0.523 (0.016) \\
ESM C & 0.426 (0.028) & 0.313 (0.044) & 0.408 (0.032) & 0.360 (0.027) & 0.526 (0.017) \\
ProGen2 M & 0.396 (0.028) & 0.291 (0.018) & 0.434 (0.015) & 0.379 (0.015) & 0.396 (0.020) \\
ProGen2 XL & 0.406 (0.021) & 0.300 (0.028) & 0.415 (0.023) & 0.384 (0.014) & 0.445 (0.012) \\
\midrule
SaProt & 0.461 (0.017) & 0.380 (0.016) & 0.488 (0.009) & 0.366 (0.020) & 0.592 (0.011) \\
ESM-3 Open & 0.432 (0.013) & 0.403 (0.013) & 0.470 (0.008) & 0.388 (0.012) & 0.641 (0.011) \\
ProSST & 0.480 (0.019) & 0.444 (0.021) & 0.532 (0.016) & 0.430 (0.018) & \textbf{0.653 (0.011)} \\
\midrule
MSA Transformer & 0.477 (0.009) & 0.324 (0.036) & 0.447 (0.013) & 0.416 (0.017) & 0.492 (0.011) \\
TranceptEVE L & 0.490 (0.014) & 0.371 (0.018) & 0.459 (0.012) & 0.458 (0.007) & 0.501 (0.011) \\
GEMME & 0.485 (0.008) & 0.380 (0.037) & 0.440 (0.015) & 0.450 (0.008) & 0.519 (0.012) \\
PoET-1 & 0.498 (0.006) & 0.391 (0.019) & 0.466 (0.006) & 0.474 (0.006) & 0.519 (0.005) \\
\midrule
S3F-MSA & 0.506 (0.008) & 0.437 (0.025) & 0.480 (0.013) & 0.476 (0.007) & 0.582 (0.008) \\
VenusREM & 0.499 (0.016) & 0.452 (0.018) & 0.535 (0.014) & 0.459 (0.016) & 0.651 (0.011) \\
PoET-2 & 0.508 (0.000) & 0.423 (0.000) & 0.503 (0.000) & 0.482 (0.000) & 0.582 (0.000) \\
\cdashline{1-6}
PoET-2 + VenusREM & \textbf{0.538 (0.010)} & \textbf{0.475 (0.013)} & \textbf{0.552 (0.010)} & \textbf{0.505 (0.010)} & 0.644 (0.007) \\
\bottomrule
\end{tabular}
}
\end{table}

\begin{table}[htbp]
\caption{Performance (Spearman $\rho$) on zero-shot DMS substitutions benchmark broken down by MSA depth. Standard error of difference to PoET-2 in parentheses.}
\label{tab:dms_subs_msa_depth}
\centering
\begin{tabular}{lccccc}
\toprule
\multirow{2}{*}{\textbf{Model}} & \multicolumn{3}{c}{\textbf{Substitutions By MSA Depth}} \\
\cmidrule(lr){2-4}
& \textbf{Low} & \textbf{Medium} & \textbf{High} \\
\midrule
ESM-2 & 0.340 (0.038)  & 0.410 (0.018) & 0.513 (0.013) \\
ESM C & 0.338 (0.040) & 0.401 (0.020) & 0.519 (0.011) \\
ProGen2 M & 0.305 (0.031) & 0.390 (0.016) & 0.422 (0.016) \\
ProGen2 XL & 0.322 (0.024) & 0.411 (0.011) & 0.442 (0.013) \\
\midrule
SaProt & 0.397 (0.027) & 0.446 (0.014) & 0.546 (0.011) \\
ESM-3 Open & 0.402 (0.017) & 0.465 (0.011) & 0.575 (0.011) \\
ProSST & 0.468 (0.029) & 0.506 (0.013) & 0.581 (0.013) \\
\midrule
MSA Transformer & 0.375 (0.024) & 0.456 (0.011) & 0.480 (0.012) \\
TranceptEVE L & 0.434 (0.015) & 0.473 (0.008) & 0.491 (0.009) \\
GEMME & 0.445 (0.017) & 0.474 (0.008) & 0.494 (0.009) \\
PoET-1 & 0.479 (0.008) & 0.477 (0.006) & 0.511 (0.005) \\
\midrule
S3F-MSA & 0.470 (0.017) & 0.509 (0.005) & 0.547 (0.007) \\
VenusREM & 0.498 (0.023) & 0.524 (0.011) & 0.578 (0.013) \\
PoET-2 & 0.488 (0.000) & 0.507 (0.000) & 0.555 (0.000) \\
\cdashline{1-4}
PoET-2 + VenusREM & \textbf{0.528 (0.016)} & \textbf{0.550 (0.007)} & \textbf{0.593 (0.008)} \\
\bottomrule
\end{tabular}
\end{table}

\begin{table}[htbp]
\caption{Performance (Spearman $\rho$) on zero-shot DMS substitutions benchmark broken down by taxonomy. Standard error of difference to PoET-2 in parentheses.}
\label{tab:dms_subs_taxon}
\centering
\begin{tabular}{lccccc}
\toprule
\multirow{2}{*}{\textbf{Model}} & \multicolumn{4}{c}{\textbf{Substitutions By Taxonomy}} \\
\cmidrule(lr){2-5}
& \textbf{Human} & \textbf{Other Eukaryote} & \textbf{Prokaryote} & \textbf{Virus} \\
\midrule
ESM-2 & 0.457 (0.011) & 0.488 (0.031) & 0.459 (0.019) & 0.262 (0.043) \\
ESM C & 0.467 (0.010) & 0.482 (0.030) & 0.442 (0.022) & 0.245 (0.051) \\
ProGen2 M & 0.412 (0.011) & 0.418 (0.027) & 0.355 (0.027) & 0.334 (0.035) \\
ProGen2 XL & 0.385 (0.012) & 0.459 (0.017) & 0.417 (0.017) & 0.401 (0.024) \\
\midrule
SaProt & 0.478 (0.010) & 0.530 (0.018) & 0.515 (0.013) & 0.323 (0.037) \\
ESM-3 Open & 0.480 (0.008) & 0.549 (0.015) & 0.530 (0.016) & 0.407 (0.028) \\
ProSST & 0.518 (0.013) & 0.577 (0.019) & 0.550 (0.018) & 0.449 (0.028) \\
\midrule
MSA Transformer & 0.439 (0.012) & 0.517 (0.012) & 0.445 (0.014) & 0.419 (0.028) \\
TranceptEVE L & 0.472 (0.007) & 0.515 (0.014) & 0.455 (0.013) & 0.460 (0.020) \\
GEMME & 0.468 (0.009) & 0.519 (0.013) & 0.467 (0.011) & 0.471 (0.019) \\
PoET-1 & 0.481 (0.005) & 0.543 (0.009) & 0.464 (0.008) & 0.491 (0.011) \\
\midrule
S3F-MSA & 0.501 (0.008) & 0.561 (0.010) & 0.521 (0.008) & 0.502 (0.012) \\
VenusREM & 0.530 (0.011) & 0.586 (0.017) & 0.550 (0.016) & 0.489 (0.023) \\
PoET-2 & 0.506 (0.000) & 0.569 (0.000) & 0.507 (0.000) & 0.528 (0.000) \\
\cdashline{1-5}
PoET-2 + VenusREM & \textbf{0.548 (0.008)} & \textbf{0.604 (0.011)} & \textbf{0.562 (0.009)} & \textbf{0.551 (0.013)} \\
\bottomrule
\end{tabular}
\end{table}

\begin{table}[htbp]
\caption{Performance (Spearman $\rho$) on zero-shot DMS substitutions benchmark broken down by mutation depth. Standard error of difference to PoET-2 in parentheses.}
\label{tab:dms_subs_mutation_depth}
\centering
\resizebox{\textwidth}{!}{
\begin{tabular}{lccccc}
\toprule
\multirow{2}{*}{\textbf{Model}} & \multicolumn{5}{c}{\textbf{Substitutions By Mutation Depth}} \\
\cmidrule(lr){2-6}
& \textbf{1} & \textbf{2} & \textbf{3} & \textbf{4} & \textbf{5+} \\
\midrule
ESM-2 & 0.423 (0.011) & 0.248 (0.021) & 0.203 (0.077) & 0.160 (0.077) & 0.220 (0.073) \\
ESM C & 0.416 (0.012) & 0.257 (0.022) & 0.189 (0.073) & 0.150 (0.073) & 0.217 (0.074) \\
ProGen2 M & 0.372 (0.010) & 0.131 (0.025) & 0.149 (0.059) & 0.131 (0.066) & 0.178 (0.062) \\
ProGen2 XL & 0.384 (0.008) & 0.181 (0.023) & 0.267 (0.046) & 0.229 (0.046) & 0.283 (0.047) \\
\midrule
SaProt & 0.459 (0.010) & 0.312 (0.018) & 0.271 (0.048) & 0.268 (0.051) & 0.337 (0.056) \\
ESM-3 Open & 0.487 (0.009) & 0.336 (0.019) & 0.303 (0.047) & 0.284 (0.046) & 0.365 (0.050) \\
ProSST & 0.520 (0.009) & 0.393 (0.025) & 0.316 (0.046) & 0.274 (0.049) & 0.334 (0.060) \\
\midrule
MSA Transformer & 0.427 (0.008) & 0.220 (0.019) & 0.358 (0.026) & 0.365 (0.017) & 0.401 (0.022) \\
TranceptEVE L & 0.446 (0.006) & 0.277 (0.014) & 0.349 (0.041) & 0.327 (0.039) & 0.385 (0.046) \\
GEMME & 0.447 (0.006) & 0.275 (0.017) & 0.329 (0.044) & 0.338 (0.028) & 0.419 (0.022) \\
PoET-1 & 0.466 (0.004) & 0.298 (0.010) & 0.412 (0.019) & 0.393 (0.014) & 0.421 (0.011) \\
\midrule
S3F-MSA & 0.499 (0.005) & 0.332 (0.011) & 0.377 (0.017) & 0.343 (0.017) & 0.387 (0.033) \\
VenusREM & 0.534 (0.008) & 0.395 (0.023) & 0.352 (0.046) & 0.320 (0.045) & 0.372 (0.050) \\
PoET-2 & 0.506 (0.000) & 0.357 (0.000) & \textbf{0.444 (0.000)} & \textbf{0.419 (0.000)} & \textbf{0.447 (0.000)} \\
\cdashline{1-6}
PoET-2 + VenusREM & \textbf{0.556 (0.005)} & \textbf{0.402 (0.014)} & 0.442 (0.023) & 0.411 (0.020) & 0.441 (0.028) \\
\bottomrule
\end{tabular}
}
\end{table}

\subsection{Compute requirements}\label{appendix:zero_shot_compute_requirements}

Inference with PoET-2 is performed on g5.xlarge instances from Amazon Web Services. The instances are equipped with A10G Nvidia GPUs with 24GB VRAM. For scoring sequences of average length (\textasciitilde $350$ amino acids), the inference throughput per prompt is approximately 125 sequences per second.

\clearpage
\section{Supervised variant effect prediction}\label{appendix:supervised}

\paragraph{Overview} As described in the main text, our supervised variant effect prediction methodology employs a Gaussian Process (GP) regression model to predict fitness scores. The GP is configured with a constant mean function and a product kernel. This kernel integrates information from two Matérn 5/2 sub-kernels, each operating on distinct features derived from PoET-2.

One sub-kernel operates on protein embeddings derived from the last layer of PoET-2's MLM decoder. Given the high dimensionality of the full per-residue embeddings produced by PoET-2, a dimensionality reduction step is applied prior to their use in the GP. Specifically, we utilize Singular Value Decomposition (SVD) to project the full embeddings into a 1024-dimensional space. This dimensionality reduction strategy is conceptually similar to the PCA-based approach used by Bepler et al. \cite{bepler_overview} to improve the runtime of their learning algorithm, which is also a GP for protein fitness prediction. For each wild-type (WT) protein in an assay, the SVD transformation is fitted on a set of 1536 variants: this set comprises the WT sequence itself and a random sample of 1535 single and double substitution mutants of that WT. The number of variants used for fitting SVDs (1536) was chosen to be approximately 50\% larger than the number of SVD components (1024). This was deemed a practical trade-off to provide a reasonable basis for the decomposition while managing the computational requirements of fitting SVD transformations for each of the numerous proteins in the ProteinGym benchmark; fitting the SVD on a larger or more diverse set of variants may improve performance.

The second Matérn 5/2 sub-kernel in the product utilizes the log likelihood ratios (LLRs) obtained from PoET-2's CLM decoder, as used in zero-shot prediction (\S\ref{zero_shot_methods}). The final GP model thus learns from both the reduced-dimensionality MLM embeddings and the CLM-derived LLRs.

\paragraph{Gaussian Process Hyperparameter Priors} To improve the stability and performance of GP training, particularly with small training datasets, we incorporate empirical priors on the GP hyperparameters. This process involves three steps: (1) We first fit individual GP models (with the architecture described above) to each of the 8 validation datasets specified by ProteinNPT \cite{proteinnpt} (listed in Appendix \ref{appendix:zero_shot_ensembles}). (2) From these 8 trained GPs, we extract the optimized hyperparameters and fit empirical distributions to them. Specifically, a Normal distribution is fitted to the learned constant mean values, while Gamma distributions are fitted to the other hyperparameters (i.e. the lengthscales of both Matérn kernels, the outputscale of the product kernel, and the likelihood noise term). (3) These fitted Normal and Gamma distributions then serve as priors for the respective hyperparameters when training GPs on the ProteinGym benchmark assays.

This prior-informed approach is particularly beneficial for assays with limited training data (e.g. fewer than \textasciitilde50 data points), where the prior helps guide the optimization process. For larger training set sizes, the influence of the prior diminishes. When conducting ablation studies involving different foundation models to generate the input embeddings and LLRs, the procedure for deriving these priors (steps 1 and 2) is repeated to learn a separate, appropriate set of hyperparameter priors for each distinct foundation model.

\subsection{Prompt engineering}\label{appendix:prompt_engineering_supervised}

Similar to our approach to prompt engineering for zero-shot variant effect prediction (Appendix \ref{appendix:prompt_engineering_zero_shot}), for supervised prediction, we also explore two prompt engineering methods.

\begin{figure}[htbp]
  \centering
  \includegraphics[width=\linewidth]{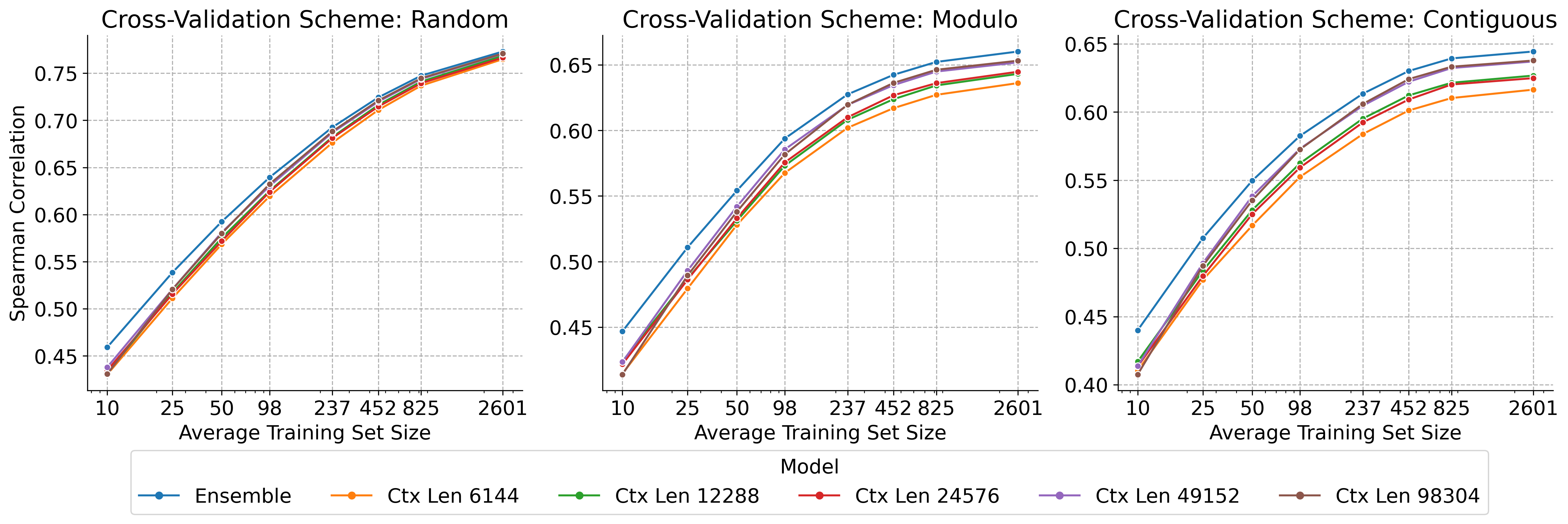}
  \caption{Performance on the supervised DMS substitutions benchmark as a function of training set size of (1) PoET-2 GP models using prompts with 5 different context lengths and (2) the ensemble GP model that ensembles the five models from (1).}
  \label{fig:data_titration_by_ctx_len}
\end{figure}

\paragraph{Ensembling over context length and maximum similarity} We ensemble over different values of context length $\in\{6144,12288,24576,49152,98304\}$ and always use a maximum similarity value of $0.95$, resulting in 5 combinations in total. We use a wide range of context lengths compared to zero-shot prediction because we observed in early experiments on ProteinNPT's validation set of 8 datasets (Appendix \ref{appendix:zero_shot_ensembles}) that longer contexts lengths generally had a small positive impact on supervised prediction performance (in contrast, for PoET-1, it was observed that long context lengths could have a negative effect on zero-shot prediction \cite{poet1}). Figure \ref{fig:data_titration_by_ctx_len} shows the performance of the GP model with different values for the context length, and the performance of the ensemble model. We use a fixed value of $0.95$ for maximum similarity because $0.95$ is typically the best value for zero-shot prediction, and did not explore other values in order to conserve compute. Therefore, it is most likely the case that there exists more optimal parameters for ensembling.

\paragraph{Incorporating structure in the prompt} Table \ref{tab:supervised_prompt_engineering} shows the performance of various strategies for incorporating structure in the prompt on an extended validation set of 30 assays from ProteinGym's DMS supervised substitutions benchmark. We find that all strategies, both those that do and don't incorporate structure in the prompt, perform about the same for supervised variant effect prediction, despite significant improvements for zero-shot variant effect prediction when using strategies that incorporate structure in the prompt. Therefore, for supervised variant effect prediction, we do not include structure in the prompt.

\begin{table}[htbp]
\centering
\caption{Performance (Spearman's $\rho$) of different strategies for including structure in the prompt on a validation set of 30 datasets from the supervised DMS substitutions benchmark.}
\label{tab:supervised_prompt_engineering}
\resizebox{\textwidth}{!}{
\begin{tabular}{ccccccc}
\toprule
\multicolumn{2}{c}{\textbf{Prompt}} & \multicolumn{4}{c}{\textbf{Supervised Cross-Validation Scheme}} & \multirow{2}{*}{\textbf{Zero-shot}} \\
\cmidrule(lr){1-2} \cmidrule(lr){3-6}
\textbf{Context Modalities} & \textbf{Query} & \textbf{Rand.} & \textbf{Mod.} & \textbf{Contig.} & \textbf{Avg.} & \\
\midrule
Sequence & None & \textbf{0.70963} & \textbf{0.60398} & \textbf{0.49490} & \textbf{0.60284} & 0.40711 \\
Sequence and Structure & None & 0.70689 & 0.59785 & 0.49312 & 0.59929 & 0.42554 \\
Sequence & Structure of WT & 0.70436 & 0.60179 & 0.49142 & 0.59919 & 0.42277 \\
Sequence and Structure & Structure of WT & 0.70103 & 0.58561 & 0.47449 & 0.58704 & \textbf{0.43748} \\
\bottomrule
\end{tabular}
}
\end{table}

The extended validation set consists of ProteinNPT's 8 validation datasets (Appendix \ref{appendix:zero_shot_ensembles}), plus 22 randomly selected datasets. The full list of extended validation datasets is as follows:

\texttt{\detokenize{BLAT_ECOLX_Jacquier_2013, CALM1_HUMAN_Weile_2017, DYR_ECOLI_Thompson_2019, DLG4_RAT_McLaughlin_2012, P53_HUMAN_Giacomelli_2018_WT_Nutlin, REV_HV1H2_Fernandes_2016, RL40A_YEAST_Roscoe_2013, TAT_HV1BR_Fernandes_2016, ACE2_HUMAN_Chan_2020, BCHB_CHLTE_Tsuboyama_2023_2KRU, CAR11_HUMAN_Meitlis_2020_gof, CP2C9_HUMAN_Amorosi_2021_abundance, ENVZ_ECOLI_Ghose_2023, F7YBW7_MESOW_Ding_2023, GCN4_YEAST_Staller_2018, GLPA_HUMAN_Elazar_2016, HCP_LAMBD_Tsuboyama_2023_2L6Q, KCNH2_HUMAN_Kozek_2020, LYAM1_HUMAN_Elazar_2016, MBD11_ARATH_Tsuboyama_2023_6ACV, MTHR_HUMAN_Weile_2021, OBSCN_HUMAN_Tsuboyama_2023_1V1C, OPSD_HUMAN_Wan_2019, PA_I34A1_Wu_2015, PSAE_SYNP2_Tsuboyama_2023_1PSE, PTEN_HUMAN_Matreyek_2021, Q53Z42_HUMAN_McShan_2019_binding-TAPBPR, RNC_ECOLI_Weeks_2023, SPG1_STRSG_Olson_2014, TADBP_HUMAN_Bolognesi_2019}}

\subsection{Gaussian Process kernels}\label{appendix:supervised_gp_kernels}

Our primary Gaussian Process (GP) model for supervised variant effect prediction utilizes a product kernel combining two Matérn 5/2 sub-kernels: one operating on PoET-2 MLM embeddings and the other on PoET-2 CLM log likelihood ratios (LLRs). While this product kernel is effective for single-site substitutions, its performance can be suboptimal when predicting the effects of multi-mutation variants due to the behavior of the LLR-based sub-kernel under distributional shift.

Specifically, the distribution of LLRs for multi-mutation variants often differs significantly from that of single-site mutations, with multi-mutation LLRs tending towards much lower or higher values. Consequently, if a GP is trained predominantly on single-site variants and then applied to predict multi-mutation variants, the LLRs of the test set can be markedly out-of-distribution (OOD) relative to the training data.

This OOD characteristic poses a challenge for the Matérn 5/2 kernel operating on LLRs. Stationary kernels like the Matérn assume that covariance is a function of the distance between inputs; when test LLRs are far outside the training distribution's range, their covariance with the training data (as modeled by this kernel) diminishes significantly. Since our model employs a product kernel, if the LLR sub-kernel assigns low covariance to a test point, the overall covariance for that point will also be low, irrespective of the embedding-based sub-kernel. In such cases, the GP prediction tends to revert towards the prior mean, offering limited predictive power for these OOD multi-mutation variants.

We leave the exploration of more sophisticated methods for incorporating LLRs into the GP for multi-mutation contexts (e.g. using LLR transformations or non-stationary kernels) to future work. For the current work, a pragmatic approach to mitigate this issue when predicting the effects of multi-mutation variants is to utilize a GP model that relies solely on the embedding-based Matérn kernel, thereby omitting the LLR-based sub-kernel for these specific predictions. Even with this simplification, a PoET-2 based GP using only embeddings can achieve strong performance in predicting the effects of multi-mutation variants when trained on data from single-site or lower-order mutants, outperforming other state-of-the-art methods as discussed in Appendix \ref{appendix:supervised_multimutant}.

\subsection{Detailed results}\label{appendix:supervised_detailed_results}

The following tables detail results on the performance and standard error of models on the DMS supervised benchmark, broken by different metrics, cross-validation schemes, and assay and protein subgroups.

\begin{itemize}
    \item \textbf{Table \ref{tab:dms_supervised_spearman_detailed}} Performance (Spearman) broken down by cross-validation scheme, with standard error of difference to PoET-2.
    \item \textbf{Table \ref{tab:dms_supervised_mse_detailed}} Performance (MSE) broken down by cross-validation scheme, with standard error of difference to PoET-2.
    \item \textbf{Table \ref{tab:dms_supervised_assay_type}} Performance (average Spearman across cross-validation schemes) on substitutions benchmark broken down by assay type, with standard error of difference to PoET-2.
    \item \textbf{Table \ref{tab:dms_supervised_msa_depth}} Performance (average Spearman across cross-validation schemes) on substitutions benchmark broken down by MSA depth, with standard error of difference to PoET-2.
    \item \textbf{Table \ref{tab:dms_supervised_taxon}} Performance (average Spearman across cross-validation schemes) on substitutions benchmark broken down by taxonomy, with standard error of difference to PoET-2.
    \item \textbf{Table \ref{tab:dms_supervised_small_dataset_size_significance}} Performance (average Spearman across cross-validation schemes) on substitutions benchmark for the smallest training set size ($n=10$) with standard error of difference to PoET-2. This table demonstrates that PoET-2's performance advantage is statistically significant even in the extreme few-shot regime. Differences for larger dataset sizes ($n>10$) are all statistically significant with $p < 1e-5$ and are omitted for brevity.
\end{itemize}

\begin{table}[htbp]
\caption{Performance (Spearman $\rho$) on supervised DMS substitutions benchmark. Standard error of difference to PoET-2 GP in parentheses.}
\label{tab:dms_supervised_spearman_detailed}
\centering
\begin{tabular}{lcccc}
\toprule
 & \multicolumn{4}{c}{\textbf{Spearman $\rho$ ($\uparrow$)}} \\
\cmidrule{2-5}
\textbf{Model} & \textbf{Random} & \textbf{Modulo} & \textbf{Contiguous} & \textbf{Average} \\
\midrule
ProteinNPT       & 0.741 (0.003) & 0.588 (0.012) & 0.529 (0.018) & 0.619 (0.010)\\
\midrule
Kermut           & 0.746 (0.004) & 0.635 (0.008) & 0.613 (0.009) & 0.664 (0.006) \\
\midrule
ESM-2 (650 M) GP & 0.749 (0.002) & 0.573 (0.009) & 0.549 (0.010) & 0.624 (0.007) \\
ESM C GP         & 0.747 (0.004) & 0.605 (0.007) & 0.573 (0.010) & 0.642 (0.006) \\
PoET-2 GP        & \textbf{0.773 (0.000)} & \textbf{0.661 (0.000)} & \textbf{0.645 (0.000)} & \textbf{0.693 (0.000)} \\
\bottomrule
\end{tabular}
\end{table}

\begin{table}[htbp]
\caption{Performance (MSE) on supervised DMS substitutions benchmark. Standard error of difference to PoET-2 GP in parentheses.}
\label{tab:dms_supervised_mse_detailed}
\centering
\begin{tabular}{lcccc}
\toprule
 & \multicolumn{4}{c}{\textbf{MSE ($\downarrow$)}} \\
\cmidrule{2-5}
\textbf{Model} & \textbf{Random} & \textbf{Modulo} & \textbf{Contiguous} & \textbf{Average} \\
\midrule
ProteinNPT       & 0.441 (0.012) & 0.765 (0.023) & 0.856 (0.025) & 0.687 (0.015)\\
\midrule
Kermut           & 0.413 (0.004) & 0.649 (0.010) & 0.697 (0.010) & 0.586 (0.007) \\
\midrule
ESM-2 (650 M) GP & 0.404 (0.004) & 0.720 (0.011) & 0.768 (0.011) & 0.630 (0.008) \\
ESM C GP         & 0.398 (0.004) & 0.660 (0.008) & 0.716 (0.009) & 0.592 (0.007) \\
PoET-2 GP        & \textbf{0.370 (0.000)} & \textbf{0.602 (0.000)} & \textbf{0.647 (0.000)} & \textbf{0.540 (0.000)} \\
\bottomrule
\end{tabular}
\end{table}

\begin{table}[htbp]
\caption{Performance (Spearman $\rho$) on supervised DMS substitutions benchmark broken down by assay type. Standard error of difference to PoET-2 GP in parentheses.}
\label{tab:dms_supervised_assay_type}
\centering
\resizebox{\textwidth}{!}{
\begin{tabular}{lccccc}
\toprule
\multirow{2}{*}{\textbf{Model}} & \multicolumn{5}{c}{\textbf{Substitutions By Assay Type}} \\
\cmidrule(lr){2-6}
& \textbf{Activity} & \textbf{Binding} & \textbf{Expression} & \makecell[c]{\textbf{Organismal}\\\textbf{Fitness}} & \textbf{Stability} \\
\midrule
ProteinNPT & 0.590 (0.007) & 0.541 (0.045) & 0.631 (0.011) & 0.558 (0.010) & 0.776 (0.005) \\
\midrule
Kermut & 0.606 (0.007) & 0.627 (0.027) & 0.680 (0.010) & 0.584 (0.007) & 0.825 (0.004) \\
\midrule
ESM-2 (650 M) GP & 0.569 (0.014) & 0.577 (0.026) & 0.633 (0.012) & 0.545 (0.010) & 0.795 (0.006) \\
ESM C GP & 0.575 (0.015) & 0.601 (0.021) & 0.656 (0.013) & 0.550 (0.013) & 0.828 (0.006) \\
PoET-2 GP & \textbf{0.630 (0.000)} & \textbf{0.667 (0.000)} & \textbf{0.691 (0.000)} & \textbf{0.622 (0.000)} & \textbf{0.854 (0.000)} \\
\bottomrule
\end{tabular}
}
\end{table}

\begin{table}[htbp]
\caption{Performance (Spearman $\rho$) on supervised DMS substitutions benchmark broken down by MSA depth. Standard error of difference to PoET-2 GP in parentheses.}
\label{tab:dms_supervised_msa_depth}
\centering
\begin{tabular}{lccccc}
\toprule
\multirow{2}{*}{\textbf{Model}} & \multicolumn{3}{c}{\textbf{Substitutions By MSA Depth}} \\
\cmidrule(lr){2-4}
& \textbf{Low} & \textbf{Medium} & \textbf{High} \\
\midrule
ProteinNPT & 0.576 (0.016)  & 0.621 (0.010) & 0.705 (0.006) \\
\midrule
Kermut & 0.619 (0.012) & 0.658 (0.005) & 0.743 (0.005) \\
\midrule
ESM-2 (650 M) GP & 0.561 (0.019) & 0.618 (0.007) & 0.721 (0.006) \\
ESM C GP & 0.581 (0.019) & 0.627 (0.010) & 0.749 (0.005) \\
PoET-2 GP & \textbf{0.667 (0.000)} & \textbf{0.689 (0.000)} & \textbf{0.769 (0.000)} \\
\bottomrule
\end{tabular}
\end{table}

\begin{table}[htbp]
\caption{Performance (Spearman $\rho$) on supervised DMS substitutions benchmark broken down by taxonomy. Standard error of difference to PoET-2 GP in parentheses.}
\label{tab:dms_supervised_taxon}
\centering
\begin{tabular}{lccccc}
\toprule
\multirow{2}{*}{\textbf{Model}} & \multicolumn{4}{c}{\textbf{Substitutions By Taxonomy}} \\
\cmidrule(lr){2-5}
& \textbf{Human} & \textbf{Other Eukaryote} & \textbf{Prokaryote} & \textbf{Virus} \\
\midrule
ProteinNPT & 0.633 (0.007) & 0.673 (0.006) & 0.666 (0.012) & 0.602 (0.019) \\
\midrule
Kermut & 0.671 (0.006) & 0.712 (0.006) & 0.707 (0.004) & 0.628 (0.012) \\
\midrule
ESM-2 (650 M) GP & 0.649 (0.006) & 0.668 (0.016) & 0.673 (0.009) & 0.552 (0.016) \\
ESM C GP & 0.674 (0.005) & 0.682 (0.017) & 0.696 (0.007) & 0.542 (0.023) \\
PoET-2 GP & \textbf{0.696 (0.000)} & \textbf{0.738 (0.000)} & \textbf{0.736 (0.000)} & \textbf{0.690 (0.000)} \\
\bottomrule
\end{tabular}
\end{table}

\begin{table}[htbp]
\caption{Mean Spearman correlation ($\rho$) and standard error on supervised DMS substitutions benchmark when training datasets are limited to no more than $n=10$ data points. Values in parentheses are the standard error of the difference in mean performance relative to PoET-2 GP, computed via 10,000 bootstrap samples. All differences are statistically significant ($p < 4.5e-3$).}\label{tab:dms_supervised_small_dataset_size_significance}
\label{tab:supervised_n10_significance}
\centering
\begin{tabular}{lccc}
\toprule
 & \multicolumn{3}{c}{\textbf{Spearman $\rho$ ($\uparrow$)}} \\
\cmidrule(lr){2-4}
\textbf{Model} & \textbf{Random} & \textbf{Modulo} & \textbf{Contiguous} \\
\midrule
ESM-2 (650M) GP  & 0.408 (0.008) & 0.389 (0.010) & 0.382 (0.012) \\
ESM-2 (3B) GP    & 0.381 (0.008) & 0.354 (0.009) & 0.354 (0.009) \\
ESM C (300M) GP  & 0.418 (0.008) & 0.399 (0.009) & 0.397 (0.011) \\
\midrule
PoET-1 GP        & 0.440 (0.004) & 0.420 (0.008) & 0.416 (0.008) \\
\midrule
PoET-2 GP        & \textbf{0.459 (0.000)} & \textbf{0.447 (0.000)} & \textbf{0.440 (0.000)} \\
\bottomrule
\end{tabular}
\end{table}

\newpage
\subsection{Comparison of MLM and CLM decoder embeddings}
\label{appendix:dms_supervised_ablation}

Table \ref{tab:dms_supervised_decoder_ablation} compares the performance of GP models trained on embeddings from PoET-2's bidirectional (MLM) decoder versus its autoregressive (CLM) decoder. We find that the MLM decoder embeddings consistently outperform the CLM decoder embeddings.

\begin{table}[htbp]
\caption{Performance of embeddings from PoET-2's MLM and CLM decoders on ProteinGym's supervised DMS substitutions benchmark.}
\label{tab:dms_supervised_decoder_ablation}
\centering
\begin{tabular}{lcccccccc}
\toprule
 & \multicolumn{4}{c}{\textbf{Spearman $\rho$ ($\uparrow$)}} & \multicolumn{4}{c}{\textbf{Mean Square Error ($\downarrow$)}} \\
\cmidrule(lr){2-5} \cmidrule(lr){6-9}
\textbf{Model} & \textbf{Rand.} & \textbf{Mod.} & \textbf{Contig.} & \textbf{Avg.} & \textbf{Rand.} & \textbf{Mod.} & \textbf{Contig.} & \textbf{Avg.} \\
\midrule
PoET-2 CLM GP & 0.757 & 0.622 & 0.601 & 0.660 & 0.398 & 0.660 & 0.713 & 0.590 \\
\textbf{PoET-2 MLM GP} & \textbf{0.771} & \textbf{0.652} & \textbf{0.637} & \textbf{0.687} & \textbf{0.374} & \textbf{0.616} & \textbf{0.657} & \textbf{0.549} \\
\bottomrule
\end{tabular}
\end{table}

\subsection{Compute requirements}\label{appendix:supervised_compute_requirements}

\begin{itemize}
    \item The computation of the SVD of embeddings from protein foundation models is performed on r6a.4xlarge instances from Amazon Web Services. These instances are equipped with 16 vCPUs and 128GB of RAM. The amount of RAM required to fit the SVD depends on the number of training samples, the length of the WT sequence, and the embedding dimension of the foundation model used. Generally we use 1536 training samples, as described at the start of this section. For some long sequences and models with very large embedding dimension, the number of samples may need to be decreased to fit within the available RAM.
    \item Embeddings and log likelihood ratios for supervised variant effect prediction are computed using the same computational resources as log likelihood ratios for zero-shot variant effect prediction (Appendix \ref{appendix:zero_shot_compute_requirements})
    \item Gaussian process models are trained on g5.xlarge instances from Amazon Web Services. These instances are equipped with A10G Nvidia GPUs that have 24GB of VRAM.
\end{itemize}

\subsection{Prediction of mutational effects of multi-mutation variants}\label{appendix:supervised_multimutant}

As discussed in Appendix \ref{appendix:supervised_gp_kernels}, when predicting multi-mutation variant effects from data on single-site or lower-order mutants, we utilize a Gaussian Process (GP) model with a kernel based solely on PoET-2 embeddings, omitting log-likelihood ratios. To evaluate this embedding-only PoET-2 GP in such challenging generalization scenarios, we benchmark it on the multi-mutation dataset introduced in the ProGen3 publication \cite{progen3}. The ProGen3 authors report that their model outperforms Kermut \cite{kermut} and matches ConFit \cite{confit} on this specific benchmark. Our PoET-2 based GP, however, surpasses all three aforementioned models (Table \ref{tab:supervised_multimutantion}).

The ProGen3 multi-mutation benchmark comprises 8 datasets selected from ProteinGym. The selection criteria, as stated by the ProGen3 authors, is as follows: "We identify all assays in ProteinGym with at least 3 mutations, and we train on all variants at most k mutations from the wild type, where k is the smallest number required for the train split to exceed 500 sequences. To ensure that the train and test splits contain proteins of similar fitness, we require that the total variation distance between the train and test distributions of fitness scores be less than 1. These filters yield 8 assays that measure diverse functional attributes for a wide range of proteins."

Table \ref{tab:supervised_multimutantion} details the performance of PoET-2 GP alongside ProGen3, Kermut, and ConFit on this multi-mutation benchmark; PoET-2 GP outperforms all other models.

\begin{table}[htbp]
\centering
\caption{Performance comparison on multi-mutation variant effect prediction benchmark.}
\label{tab:supervised_multimutantion}
\begin{tabular}{lc}
\toprule
\textbf{Model} & \textbf{Spearman $\rho$ ($\uparrow$)} \\
\midrule
Kermut         & 0.628 \\
ConFit         & 0.679 \\
ProGen3        & 0.673 \\
PoET-2 GP      & \textbf{0.708} \\
\bottomrule
\end{tabular}
\end{table}

\clearpage
\section{Licenses}\label{appendix:licenses}

Existing assets used in this paper are licensed as follows:

\begin{itemize}
    \item ProteinGym benchmark: MIT license
    \item UniRef protein database: CC BY 4.0 license
    \item AlphaFold database: CC BY 4.0 license
    \item ESM C \cite{esmc} model: Cambrian Open License Agreement \cite{cambrian_open_license}
\end{itemize}

\section{Additional details}

\subsection{Statistical significance analysis}
Statistical significance for all experiments is assessed by performing a non-parametric, two-sided bootstrap test with at least 10,000 samples. Bootstrap is performed using the same methodology as in ProteinGym \cite{proteingym}.

\section{ProteinGym Assay Dataset Sources}

We thank the authors of the original publications from which the ProteinGym assays were derived for making their experimental data publicly available \cite{sourisseau_deep_2019,deng_deep_2012,amorosi_massively_2021,chiasson_multiplexed_2020,wrenbeck_single-mutation_2017,brenan_phenotypic_2016,kozek_high-throughput_2020,davidi_highly_2020,sarkisyan_local_2016,olson_comprehensive_2014,duenas-decamp_saturation_2016,thompson_altered_2020,starita_activity-enhancing_2013,araya_fundamental_2012,wu_high-throughput_2014,young_deep_2021,jiang_balance_2016,melamed_deep_2013,jia_massively_2021,mavor_determination_2016,glazer_deep_2020,kelsic_rna_2016,rockah-shmuel_systematic_2015,kitzman_massively_2015,aakre_evolving_2015,bolognesi_mutational_2019,flynn_comprehensive_2022,haddox_mapping_2018,stiffler_evolvability_2015,tripathi_molecular_2016,findlay_accurate_2018,lee_deep_2018,weile_framework_2017,qi_quantitative_2014,chan_correlation_2017,melnikov_comprehensive_2014,nutschel_systematically_2020,jacquier_capturing_2013,mishra_systematic_2016,pokusaeva_experimental_2019,fernandes_functional_2016,sinai_generative_2021,wu_functional_2015,matreyek_integrating_2021,mighell_saturation_2018,russ_evolution-based_2020,gonzalez_fitness_2019,macdonald_dimple_2023,jones_structural_2020,chen_comprehensive_2020,kennouche_deep_2019,haddox_experimental_2016,roscoe_systematic_2014,klesmith_comprehensive_2015,gonzalez_somermeyer_heterogeneity_2022,doud_accurate_2016,soh_comprehensive_2019,seuma_genetic_2021,dandage_differential_2018,firnberg_comprehensive_2014,adkar_protein_2012,giacomelli_mutational_2018,suiter_massively_2020,hobbs_saturation_2022,gersing_comprehensive_2022,staller_high-throughput_2018,bandaru_deconstruction_2017,bridgford_novel_2020,spencer_deep_2017,mclaughlin_jr_spatial_2012,doud_site-specific_2015,mattenberger_globally_2021,matreyek_multiplex_2018,flynn_comprehensive_2020,ahler_combined_2019,romero_dissecting_2015,faure_mapping_2022,roscoe_analyses_2013,newberry_robust_2020,wu_adaptation_2016,starr_deep_2020,brauer_comprehensive_2021,linsky_novo_2020,stadelmann_deep_2021,koch_optimization_2022,kotler_systematic_2018,campbell_determinants_2022,ding_co-evolution_2022,tsuboyama_mega-scale_2023,suphatrakul_functional_2023,weile_shifting_2021,xie_predicting_2023,roychowdhury_microfluidic_2022,andrews_distinct_2020,lo_functional_2023,gajula_high_2014,chakraborty_profiling_2021,weng_energetic_2022,ding_protein_2023,nguyen_molecular_2023,seuma_atlas_2022,ursu_massively_2022,wrenbeck_automated_2019,hom_deep_2019,gray_elucidating_2019,veerapandian_directed_2018,thyagarajan_inherent_2014,ostermaier_functional_2014,bloom_experimentally_2014,hietpas_experimental_2011,gill_multiple_2023,meier_deep_2023,tan_high_2023,macrae_protein_2023,ghose_marginal_2023,nguyen_genetic_2023,clausen_mutational_2023,vanella_understanding_2023,lei_mutational_2023,loggerenberg_systematically_2023,weeks_fitness_2023,muhammad_high_2023,yee_full_2023,chen_deep_2023,gersing_characterizing_2023,huttinger_deep_2021,kwon_structurefunction_2022,sun_proactive_2020,wan_characterizing_2019,chan_engineering_2020,silverstein_systematic_2021,zinkus-boltz_phage-assisted_2019,klesmith_retargeting_2019,elazar_mutational_2016,coyote-maestas_determinants_2022,li_deep_2023,meitlis_multiplexed_2020,uk_monogenic_diabetes_consortium_prospective_2016,estevam_conserved_2023,miller_allosteric_2022,erwood_saturation_2022,jiang_exhaustive_2019,chen_hotprotein_2022}.

\newpage
\section*{NeurIPS Paper Checklist}

\begin{enumerate}

\item {\bf Claims}
    \item[] Question: Do the main claims made in the abstract and introduction accurately reflect the paper's contributions and scope?
    \item[] Answer: \answerYes{} 
    \item[] Justification: The main claims of the paper are stated in the abstract and introduction. The claims are supported by experimental evidence as discussed in Section \ref{experiments}.

\item {\bf Limitations}
    \item[] Question: Does the paper discuss the limitations of the work performed by the authors?
    \item[] Answer: \answerYes{} 
    \item[] Justification: The limitations of the paper are discussed in Section \ref{conclusion_and_limitations}.

\item {\bf Theory assumptions and proofs}
    \item[] Question: For each theoretical result, does the paper provide the full set of assumptions and a complete (and correct) proof?
    \item[] Answer: \answerNA{} 
    \item[] Justification: The paper does not include theoretical results.

    \item {\bf Experimental result reproducibility}
    \item[] Question: Does the paper fully disclose all the information needed to reproduce the main experimental results of the paper to the extent that it affects the main claims and/or conclusions of the paper (regardless of whether the code and data are provided or not)?
    \item[] Answer: \answerYes{} 
    \item[] Justification: Pre-training details are provided in Appendix \ref{appendix:poet_2_training_details}. Training and evaluation details of downstream models are described in Section \ref{experiments}, Appendix \ref{appendix:zero_shot}, and Appendix \ref{appendix:supervised}.

\item {\bf Open access to data and code}
    \item[] Question: Does the paper provide open access to the data and code, with sufficient instructions to faithfully reproduce the main experimental results, as described in supplemental material?
    \item[] Answer: \answerNo{} 
    \item[] Justification: Code and model weights are planned for future public release.

\item {\bf Experimental setting/details}
    \item[] Question: Does the paper specify all the training and test details (e.g., data splits, hyperparameters, how they were chosen, type of optimizer, etc.) necessary to understand the results?
    \item[] Answer: \answerYes{} 
    \item[] Justification: The experimental setting is described in Section \ref{experiments}.

\item {\bf Experiment statistical significance}
    \item[] Question: Does the paper report error bars suitably and correctly defined or other appropriate information about the statistical significance of the experiments?
    \item[] Answer: \answerYes{} 
    \item[] Justification: Statistical significance of results relating to claims are provided in Section \ref{experiments}.

\item {\bf Experiments compute resources}
    \item[] Question: For each experiment, does the paper provide sufficient information on the computer resources (type of compute workers, memory, time of execution) needed to reproduce the experiments?
    \item[] Answer: \answerYes{} 
    \item[] Justification: Information about compute resources required to reproduce experiments are provided in Appendix \ref{appendix:poet_2_training_details}, Appendix \ref{appendix:zero_shot_compute_requirements}, and Appendix \ref{appendix:supervised_compute_requirements}.
    
\item {\bf Code of ethics}
    \item[] Question: Does the research conducted in the paper conform, in every respect, with the NeurIPS Code of Ethics \url{https://neurips.cc/public/EthicsGuidelines}?
    \item[] Answer: \answerYes{} 
    \item[] Justification: The research conform with the NeurIPS Code of Ethics.

\item {\bf Broader impacts}
    \item[] Question: Does the paper discuss both potential positive societal impacts and negative societal impacts of the work performed?
    \item[] Answer: \answerYes{} 
    \item[] Justification: The broader impacts are discussed in Appendix \ref{appendix:broader_impacts}.
    
\item {\bf Safeguards}
    \item[] Question: Does the paper describe safeguards that have been put in place for responsible release of data or models that have a high risk for misuse (e.g., pretrained language models, image generators, or scraped datasets)?
    \item[] Answer: \answerNA{} 
    \item[] Justification: The paper poses no such risks.

\item {\bf Licenses for existing assets}
    \item[] Question: Are the creators or original owners of assets (e.g., code, data, models), used in the paper, properly credited and are the license and terms of use explicitly mentioned and properly respected?
    \item[] Answer: \answerYes{} 
    \item[] Justification: Papers that produced existing assets used in this paper are cited. The licenses are specified in Appendix \ref{appendix:licenses}.

\item {\bf New assets}
    \item[] Question: Are new assets introduced in the paper well documented and is the documentation provided alongside the assets?
    \item[] Answer: \answerNA{} 
    \item[] Justification: The paper does not release new assets.

\item {\bf Crowdsourcing and research with human subjects}
    \item[] Question: For crowdsourcing experiments and research with human subjects, does the paper include the full text of instructions given to participants and screenshots, if applicable, as well as details about compensation (if any)? 
    \item[] Answer: \answerNA{} 
    \item[] Justification: The paper does not involve crowdsourcing nor research with human subjects.

\item {\bf Institutional review board (IRB) approvals or equivalent for research with human subjects}
    \item[] Question: Does the paper describe potential risks incurred by study participants, whether such risks were disclosed to the subjects, and whether Institutional Review Board (IRB) approvals (or an equivalent approval/review based on the requirements of your country or institution) were obtained?
    \item[] Answer: \answerNA{} 
    \item[] Justification: The paper does not involve crowdsourcing nor research with human subjects.

\item {\bf Declaration of LLM usage}
    \item[] Question: Does the paper describe the usage of LLMs if it is an important, original, or non-standard component of the core methods in this research? Note that if the LLM is used only for writing, editing, or formatting purposes and does not impact the core methodology, scientific rigorousness, or originality of the research, declaration is not required.
    \item[] Answer: \answerNA{} 
    \item[] Justification: LLMs are not used for the development of core methods in this research.

\end{enumerate}
\end{document}